\documentclass[9pt,journal]{IEEEtran}

\usepackage[dvips,final]{graphicx}
\usepackage[dvips]{color}

\usepackage{amsmath,amsfonts,amsthm,amssymb}
\usepackage[caption=false,font=footnotesize]{subfig}

\usepackage{mathtools}

\usepackage{epsfig,bm}
\usepackage{graphicx}

\usepackage[switch]{lineno}
\usepackage[named]{algo}
\usepackage[noend]{algpseudocode}
\usepackage{algorithm}

\usepackage[normalem]{ulem}
\usepackage{xcolor}
\newcommand\redsout{\bgroup\markoverwith{\textcolor{red}{\rule[0.5ex]{2pt}{0.4pt}}}\ULon}
\usepackage{bm}
\usepackage{booktabs}
\usepackage{stfloats}

\usepackage{dsfont}
\usepackage{esint}

\usepackage{url}
\usepackage{balance}

\usepackage{cite}

\newcommand{\beq}{\begin{equation}}
\newcommand{\eeq}{\end{equation}}
\newcommand{\beqn}{\begin{eqnarray}}
\newcommand{\eeqn}{\end{eqnarray}}

\newcommand\atopnew[2]{\genfrac{}{}{0pt}{}{#1}{#2}}

\newtheorem{theorem}{Theorem}

\newtheorem{proposition}[theorem]{Proposition}

\begin{document}

\markboth{IEEE Transactions on Radar Systems}%
{SUN \MakeLowercase{\textit{et al.}}: Widely Separated MIMO Radar Using Matrix Completion}

\title{Widely Separated MIMO Radar Using Matrix Completion}

\author{Shunqiao~Sun, \emph{Senior Member, IEEE}, Yunqiao~Hu, \emph{Student Member, IEEE},
		Kumar Vijay~Mishra, \emph{Senior Member, IEEE},  
        and Athina P.~Petropulu,  \emph{Fellow, IEEE} 
\thanks{S. Sun and Y. Hu are with the Department of Electrical and Computer Engineering, The University of Alabama, Tuscaloosa, AL 35487 USA. E-mails: shunqiao.sun@ua.edu, yhu62@crimson.ua.edu.}
\thanks{K. V. Mishra is with the United States DEVCOM Army Research Laboratory, Adelphi, MD 20783 USA. E-mail: kvm@ieee.org.}%
\thanks{A. P. Petropulu is with the Department of Electrical and Computer Engineering, Rutgers, The State University of New Jersey, Piscataway, NJ 08854 USA. E-mail: athinap@rutgers.edu.}
\thanks{The conference precursor of this work was presented at the 2019 IEEE Radar Conference \cite{Sun_WS_MIMO_Radarconf_2019}.  
This work was supported in part by U.S. National Science Foundation (NSF) under Grant CCF-2153386. }
}

\maketitle

\begin{abstract}
We present a low-complexity widely separated multiple-input-multiple-output (WS-MIMO) radar that samples the signals at each of its multiple receivers at reduced rates. We process the low-rate samples of all transmit-receive chains at each receiver as data matrices. We demonstrate that each of these matrices is low rank as long as the target moves slowly within a coherent processing interval. We leverage matrix completion (MC) to recover the missing samples of each receiver signal matrix at the common fusion center. Subsequently, we estimate the targets' positions and Doppler velocities via the maximum likelihood method. Our MC-WS-MIMO approach recovers missing samples and thereafter target parameters at reduced rates without discretization. Our analysis using ambiguity functions shows that antenna geometry affects the performance of MC-WS-MIMO. Numerical experiments demonstrate reasonably accurate target localization at SNR of 20 dB and sampling rate reduction to $20$\%.
\end{abstract}


%
\IEEEpeerreviewmaketitle

\section{Introduction}
\label{sec:intro}

During the past decade, there has been extensive research interest in  multiple-input-multiple-output (MIMO) radars that employ several transmit (Tx) and receive (Rx) antennas \cite{fishler2004mimo,li2007mimo,bekkerman2006target}. 
MIMO radars are usually classified as \textit{colocated} or \textit{widely separated} depending on the antenna placement relative to the targets. In a colocated MIMO (CL-MIMO) radar \cite{li2007mimo,khan2014ambiguity}, the antennas are placed close enough to observe coherent signals reflected from a target whose radar cross-section (RCS) appears identical to all Tx-Rx antenna pairs. Unlike phased array radar 
that transmits a single waveform,  CL-MIMO transmit multiple mutually orthogonal signals. The waveform diversity can can be exploited to achieve high angular resolution \cite{godrich2010target,boyer2011performance}, and high-quality parameter identifiability 
\cite{haimovich2008mimo}. In widely separated MIMO (WS-MIMO) radar, the distance between any two antennas is much larger than their distance from the target, resulting in each Tx-Rx  antenna pair seeing a different RCS of the target. 

In this paper, we focus on WS-MIMO systems. The spatial diversity in WS-MIMO is advantageous in detecting targets with small backscatter and low speed \cite{dianat2013target,he2010target}. The angular diversity provides WS-MIMO a better probability of detection; however, this comes at a cost of increased minimum required signal-to-noise ratio (SNR), below which a phased array radar shows better detection performance \cite{fishler2006spatial}. In particular, WS-MIMO exhibits superior 
detection of Swerling chi-squared target models I and III, that are statistically independent from scan-to-scan, than the CL-MIMO \cite{aittomaki2010performance}. In an electronic warfare scenario, a WS-MIMO is capable of maintaining the same detection threshold as a monostatic radar but with a lower total radiated power. This decreases the probability of intercepting the radar's
transmit signal by hostile entities \cite{stoica2007probing}.

WS-MIMO radar are similar to 
traditional multi-static radar in the sense that they  both employ  widely separated Tx-Rx units,
but they differ from multi-static radar in the way they make a decision about the target. Multi-static radar 
perform a significant amount of local processing at each receiver, and they use central unit to fuse the  decisions of the local units.  WS-MIMO radar jointly process the signals from all the receivers and make a single decision about the target.  This joint processing approach is beneficial in detecting a spatially diverse target which requires probing from different directions \cite{dianat2013target,majd2017spatial}, or a \textit{stealth} target \cite{griffiths2010multistatic}. In the latter case, when the target becomes unobservable for specific Tx-Rx pairs, it may escape detection altogether in a multistatic radar because of local processing at each receiver \cite{mrstik1978multistatic}.

Even though a WS-MIMO radar provides superior detection over the conventional multi-static network, the use of multiple waveforms in MIMO systems implies the need for several Tx-Rx radio-frequency (RF) chains,  resulting in huge hardware cost, high energy consumption, and very large computational complexity \cite{brookner2014mimo,godrich2011power}. Further, the joint processing of all receivers requires the transmission of the measurements of each antenna to the fusion center, thus involving an additional communications cost. Lately, various techniques have been proposed to address the problem of reducing the cost of hardware, energy, and area in conventional MIMO radars (see, e.g., \cite{mishra2019sub} for a review). These methods exploit the fact that the target scene is \textit{sparse}, and the radar processing tasks can be modeled as finding sparse solutions to under-determined linear equations - an aspect addressed by the emerging field of compressed sensing (CS) \cite{baraniuk2007compressive}.

There is a large body of literature on CS-based CL-MIMO radars that focuses on processing with a reduced number of signal samples as may be the case while using fewer RF chains \cite{strohmer2014sparse,yu2011measurement,yu2011cssf}.
In comparison, relatively fewer studies have examined CS applications to WS-MIMO.  The earliest application of CS to recover direction-of-arrival (DOA) with sub-Nyquist samples in a WS-MIMO was formulated in \cite{petropulu2008distributed}. This was later extended to recovering both position and Doppler velocity of the targets by reducing only the temporal sampling rate in \cite{gogineni2011target,petropulu2011exploring,li2014efficient}; performance guarantees for recovery were provided in \cite{li2014performance}. A few other studies have exploited sparsity in WS-MIMO toward power allocation \cite{yu2012study,yu2014power}, optimal detectors \cite{wang2010compressed}, dictionary learning \cite{raja2016parametric}, and stationary target imaging \cite{raja2017through}. In nearly all of these works, the targets are assumed to be located on an angle-Doppler-range grid. In practice, target parameters are typically continuous values the discretization of which may introduce gridding errors \cite{chi2011sensitivity}.

In order to avoid the off-grid errors while also maintaining high resolution, reduced-rate sampling, and low complexity, predominantly two approaches have been proposed. The first technique \cite{heckel2016superrad} formulates the radar parameter estimation for off-grid targets using atomic norm minimization \cite{mishra2015spectral} and applies to a CL-MIMO radar. 
In the second approach, the signal samples received from an array radar are processed as data matrices, which, under certain conditions, are low rank. Then, random temporal sampling at each receiver results in a partially observed data matrix and the missing entries are retrieved using \textit{matrix completion} (MC) \cite{candes2009exact,candes2010matrix,candes2010power}. Once the matrix is recovered, conventional methods are employed for target parameter recovery and estimation. The MC-based sampling and recovery was first suggested for volumetric targets in a phased-array weather radar \cite{mishra2014compressed} and later for point targets in a CL-MIMO radar \cite{sun2013target,sun2015mimomc,kalogerias2014matrix,sun2015waveform}. In addition to avoiding the grid issue, the MC approach restores the SNR loss because of subsampling. In particular, \cite{kalogerias2014matrix} provided recovery guarantees for MC-based CL-MIMO (MC-CL-MIMO) while corresponding sampling strategies and waveforms were analyzed in \cite{sun2015mimomc}. 

In this paper, we propose off-grid target recovery using MC for a WS-MIMO. The different formulation of this problem, as compared to the CL-MIMO problem makes the extension of prior work to this scenario non-trivial. 
In MC-CL-MIMO \cite{sun2015mimomc}, the low-rank data matrix is formulated by samples from all Tx-Rx chains for a single pulse. In MC-based WS-MIMO (MC-WS-MIMO), we exploit the low-rank structure of a matrix formed by samples of a single Tx-Rx chain for all pulses.

Preliminary results of this work appeared in our conference publication \cite{Sun_WS_MIMO_Radarconf_2019}, which presented initial simulation results for the target localization using the maximum likelihood (ML) approach. In this paper, we also analyze the coherence of the WS-MIMO data matrix that guarantees recovery with MC, investigate the WS-MIMO radar ambiguity function (AF), derive the Cram\'er-Rao lower bound (CRLB) of WS-MIMO radar localization, and provide more comprehensive numerical studies with comparisons to the geometry-based approach. We show that target parameters can be estimated with reasonable accuracy at $20$ dB SNR even when the sampling rate is reduced by $20$\%. Hence, MC offers notable advantages in improving the accuracy of target localization and velocity estimation, particularly in scenarios with low signal-to-noise ratios and reduced sampling rates. Additionally, the MC-based ML estimation exhibits robustness in target localization in these settings. 
Furthermore, the analysis of AF indicates that the distribution of antennas in WS-MIMO radar impacts the MC-based recovery. In our study, circularly-placed antennas are found to have improved localization than other geometries. 

The rest of the paper is organized as follows. In the next section, we introduce the system model of WS-MIMO radar in the context of the MC problem. Section~\ref{sec:coherence} provides theoretical guarantees for the coherence and recoverability of the data matrix. In Section~\ref{sec:tgt_loc}, we present the method for estimating the target parameters such as location and velocity. We provide the AF and lower error bounds for our system in Section~\ref{sec:perf}. We validate our methods through numerical experiments in Section~\ref{sec:num_exp} and conclude in Section~\ref{sec:summ}.

Throughout this paper, we reserve boldface lowercase, boldface uppercase, and calligraphic letters for vectors, matrices, and index sets, respectively. The $i$-th element of a vector $\mathbf{y}$ is $y(i)$; the $(i,j)$-th entry of a matrix $\textbf{Y}$ is $[\textbf{Y}]_{i,j}$; the $i$-th column of matrix $\textbf{Y}$ is $\textbf{y}_{i}$; and the $i$-th row of matrix $\textbf{Y}$ is $\textbf{Y}^{(i)}$. The set of $N$-dimensional vectors of complex numbers is $\mathbb{C}^N$. 
We use $\mathbf{I}_N$ for the identity matrix of size $N \times N$. We denote the transpose, Hermitian, modulus and floor operations by $(\cdot)^T$, $(\cdot)^H$, $||\cdot||$, and $\big\lfloor . \big\rfloor$, respectively. The Hadamard (point-wise) and inner products are denoted by  $\odot$ and $\langle \cdot, \cdot \rangle$, respectively. The notations $\|\cdot\|_{\ast}$ and $\|\cdot\|_{\mathcal{F}}$ are reserved for the nuclear and Frobenius norms of the matrix, respectively. The function $\text{max}(\cdot)$ returns the maximum value of its argument. The cardinality of the set is given by $|\cdot|$.

\section{System Model}
\label{sec:sysmod}
We introduce the system model of WS-MIMO radar and show that the data matrix at each receive antenna has a low-rank structure. We then propose a reduced-rate sampling scheme at each receive antenna to obtain partially observed matrices on which we apply MC technique at a fusion center.

\subsection{WS-MIMO Radar}
Consider a WS-MIMO radar system with $M_t$ transmit and $M_r$ receive antennas, located in a two-dimensional (2-D) Cartesian coordinate system. We denote the position vectors of the $m$-th transmit and $n$-th receive antennas by ${\bf{p}}_t^{(m)}= \left[ {x_t^{(m)},y_t^{(m)}} \right]^T$ and ${\bf{p}}_r^{(n)}=\left[ {x_r^{(n)},y_r^{(n)}} \right]^T$, respectively. 
The waveform orthogonality in WS-MIMO radar is achieved through either time, code, or frequency division multiplexing (TDM, CDM, or FDM) \cite{Sun_MIMO_Waveforms_Comparison_2014}.
In this paper, we employ CDM to achieve waveform orthogonality. 
Each transmit antenna emits a narrowband phase-coded pulse, composed of $N$ subpulses, during each pulse repetition interval (PRI), $T_{\textrm{PRI}}$; its reciprocal is the pulse repetition frequency (PRF). The baseband waveform of the $m$-th antenna is \cite{he2009designing} 
\begin{align}
    {s_m}\left( t \right) = \frac{1}{{\sqrt {{T_p}} }}\sum\limits_{n = 1}^N {{x_m}\left( n \right)p\left[ {\frac{{t - \left( {n - 1} \right){t_b}}}{{{t_b}}}} \right]} ,m = 1, \cdots ,{M_t},
\end{align}
where ${x_m}\left( n \right) = {e^{j{\phi _m}\left( n \right)}},m=1,\cdots,M_t, n=1,\cdots, N$ is the phase code, and $p\left( t \right)$ is the rectangular subpulse shaping function with amplitude $1$ for duration from $0$ to $1$. Here, $t_b$ is subpulse duration, and $T_p=Nt_b$ is the pulse duration.
The orthogonality implies 
$\int_{T_p} {{s_i}\left( t \right)s_j^*\left( t \right)dt = \delta \left( {i - j} \right),\;\forall i,j}$, 
where $\delta \left(  \cdot  \right)$ is the Dirac delta function.  
The transmit waveforms are narrowband such that 
\begin{align}
\frac{\lambda }{{{c}}} \ll \frac{1}{B},
\end{align}
where $\lambda = c/f_c$ is the operating wavelength of $m$-th transmitter, $f_c$ is carrier frequency, $c=3\times 10^8$ m/s is the speed of light, and $B$ is the bandwidth of
WS-MIMO radar system. Each transmitter sends out a pulse train consisting of $Q$ uniformly spaced known pulses $s_m(t)$:
\begin{align}
\label{eq:transmit_signal}
s_{m,Q}(t) = \sum_{q=0}^{Q-1} s_m\left(t-qT_{\textrm{PRI}}\right),\quad 0 \leq t \leq QT_{\textrm{PRI}}.
\end{align}
The duration of all $Q$ pulses is known as the coherent processing interval (CPI).

Assume that the radar target scene consists of $K$ targets distributed in an area denoted by a set of coordinates $\mathcal S$, sharing the same 2-D plane as the WS-MIMO transmitters and receivers. 
The $k$-th target is represented by its gravity center \cite{haimovich2008mimo} whose position vector is denoted as ${{\bf{p}}^{(k)}} = \left[ {{x_k},{y_k}} \right]^T$ moving at a velocity of ${{\bm{\nu}}^{(k)}} = \left[ {v_x^{(k)},v_y^{(k)}} \right]^T$. 
The transmit signal is reflected back by the targets and these echoes are collected by each receive antenna. For a given spatially diverse $k$-th target and $m$-th-Tx-and-$n$-th-Rx pair, the radar processor aims to retrieve following information from the received signals: reflection coefficient  ${\beta _{mn}^{(k)}}$, wherein we assumed that the target follows the Swerling I model \cite{skolnik2008radar} so that its reflectivity is constant during the CPI; time delay $\tau _{mn}^{(k)}$, which is linearly proportional to the target's location $\bf{p}^{(k)}$ as 
\begin{align}
\tau _{mn}^{(k)} = &\frac{{\left\| {{{\bf{p}}^{(k)}} - {\bf{p}}_t^{(m)}} \right\| + \left\| {{{\bf{p}}^{(k)}} - {\bf{p}}_r^{(n)}} \right\|}}{c},\label{eq:delay}
\end{align}
and Doppler frequency $f_{{mn}}^{k}$, which is proportional to the target's radial velocity $\bm{\nu}^{(k)}$ as
\begin{align}
f_{{{mn}}}^{(k)} = & \frac{{{f_c}}}{c}\left( {\frac{{\left\langle {{{\bm{\nu}}^{(k)}},{{\bf{p}}^{(k)}} - {\bf{p}}_t^{(m)}} \right\rangle }}{{\left\| {{{\bf{p}}^{(k)}} - {\bf{p}}_t^{(m)}} \right\|}} + \frac{{\left\langle {{{\bm{\nu}}^{(k)}},{{\bf{p}}^{(k)}} - {\bf{p}}_r^{(n)}} \right\rangle }}{{\left\| {{{\bf{p}}^{(k)}} - {\bf{p}}_r^{(n)}} \right\|}}} \right). \label{eq_Doppler_shift}
\end{align}

\subsection{Operating Conditions}
We make the following assumptions on the radar operation and target parameters:
\begin{description}
\item [\textbf{C1}] ``Unambiguous time-frequency region'': The target locations are assumed to lie in the unambiguous region of delay-Doppler plane $[0,R_{\max}]\times [0,\nu_{\max}]$, where $R_{\mathrm{max}}=\frac{cT_{\rm PRI}}{2}$ is the maximum unambiguous range and $\nu_{\mathrm{max}}=\frac{c}{f_0T_{\rm PRI}}$ is the maximum unambiguous velocity in both x- and y-directions, i.e. the time delays are no longer than the PRI and Doppler frequencies are up to the PRF. 
\item [\textbf{C2}] ``Low acceleration'': The frequency modulation because of a slow-moving target manifests as a frequency shift in the received signal. The targets are slow-moving and have low acceleration so that their time delays and Doppler frequencies are assumed constant over a CPI: 
\begin{align}
\frac{{2{v_i^{(k)}}Q{T_{\rm PRI}}}}{c} & \ll \frac{1}{B},\;\;i=x,y, \\
f_{mn}^{(k)}T_p & \ll 1.
\end{align} 
\item [\textbf{C3}] ``Constant delays'': The Doppler shifts induced are small over a CPI under the condition \textbf{C2} so that the delay is approximated to be constant. This allows for the piecewise-constant approximation:  $f_{mn}^{(k)}t \approx f_{mn}^{(k)}q T_{\rm PRI}, \text{for} ~t \in [q T_{\rm PRI}, (q+1))T_{\rm PRI}]$. 
\item [\textbf{C4}] ``Constant Doppler shifts'': The velocity change of a target over a CPI is small compared with the velocity resolution such that $\frac{d{\nu}_{(\cdot)}^{(k)}}{dt} \ll \frac{c}{2f_0 (Q T_{\rm PRI})^2}$, where the subscript $(\cdot)$ denotes either x- or y-directions.
\item [\textbf{C5}] ``Constant reflectivities'': The radar-to-target distance is large compared with the displacement of the target during a CPI, allowing the attenuation to be considered constant over a CPI. 
\item [\textbf{C6}] ``Unimodular waveforms'':  Due to practical hardware limitations such as amplifiers and analog-to-digital converters, the waveforms  need to be unimodular, that is, they must maintain a constant modulus.
\end{description}

\subsection{Receive Data Matrix with Low Rank Structure}
\label{subsec:lrs}
The received signal at the $n$-th receive antenna is 
\par\noindent\small
\begin{flalign}
{y_n}\left( t \right) = &\sum\limits_{q = 0}^{Q-1}\sum\limits_{m = 1}^{{M_t}} \sum\limits_{k = 1}^K \sqrt E \beta _{mn}^{(k)}{s_m}\left( {t - \tau _{mn}^{(k)} - qT_{\textrm{PRI}}} \right) \nonumber \\
& \times {e^{\mathrm{j}2\pi \left( {{f_c} + f_{mn}^{(k)}} \right)\left( {t - \tau _{mn}^{(k)}} \right)}} + {w_n}\left( t \right),
\end{flalign}\normalsize
where ${w_n}\left( t \right)$ is the additive spatio-temporally white, zero mean Gaussian noise with variance ${\sigma}_{n}^2$. 
After demodulation and passing the baseband signal through an anti-aliasing low-pass filter, the received signal at the $n$-th receiver for the $m$-th carrier frequency is 
\par\noindent\small
\begin{align}
{y_{mn}}\left( t \right) =&\sum\limits_{q = 0}^{Q-1}\sum\limits_{k = 1}^K \sqrt E \beta _{mn}^{(k)}{s_m}\left( {t - \tau _{mn}^{(k)} - qT_{\textrm{PRI}}} \right) \nonumber \\
& \times {e^{\mathrm{j}2\pi f_{mn}^{(k)}t}}{e^{ - j2\pi \left( {{f_c} + f_{mn}^{(k)}} \right)\tau _{mn}^{(k)}}}  + {w_{mn}}\left( t \right).
\end{align}\normalsize
For the sake of simplicity, the term ${{e^{ - j2\pi \left( {{f_c} + f_{mn}^{(k)}} \right)\tau _{mn}^{(k)}}}}$ can be absorbed into the target reflection coefficient ${\beta _{mn}^{(k)}}$. Hence, the baseband received signal at the $n$-th receive antenna because of the signal transmitted from the $m$-th transmit antenna is
\par\noindent\small
\begin{align}
&{y_{mn}}\left( t \right) = \nonumber \\
&\sum\limits_{q = 0}^{Q-1}\sum\limits_{k = 1}^K {\sqrt E}{\beta _{mn}^{(k)}{s_m}\left( {t - \tau _{mn}^{(k)}  - qT_{\textrm{PRI}}} \right){e^{\mathrm{j}2\pi f_{mn}^{(k)}t}}}  + {w_{mn}}\left( t \right), \label{eq_receive_with_noise}
\end{align}\normalsize
where $w_{mn}(t)$ is the noise term.

We denote the maximum and minimum ranges of all possible target locations in the coordinate set $\mathcal {S}$ with respect to the $m$-th transmit and $n$-th receive antennas as $R_{mn}^{(\max)}$ and $R_{mn}^{(\min)}$, respectively. Denote the Nyquist sampling interval by $T_s$ so that $N$ samples of the pulse duration ($T_p=NT_s$) are obtained.
For each pulse received, we set the sampling window length at each receive antenna as $N+ L_{mn}^{(\max)}$ to collect unambiguous samples for every possible location in an area covered by $\mathcal S$, where  $L_{mn}^{(\max)} = \left\lfloor {\frac{{R_{mn}^{(\max)} - R_{mn}^{(\min)}}}{{c{T_s}}}} \right\rfloor $. 
Define $L_{mn}^{(k)} = \left\lfloor {\frac{{R_{mn}^{(k)} - R_{mn}^{(\min)}}}{{c{T_s}}}} \right\rfloor $, where $R_{mn}^{(k)} = c\tau_{mn}^{(k)}$ 
is the distance corresponding to the total time of flight from the $m$-th transmitter to the $k$-th target and back from the same target to the $n$-th receiver.

Define ${{\bf{C}}_{L_{mn}^{(k)}}} = \left[ {\begin{array}{*{20}{c}}
   {\bf{0}}_{N \times L_{mn}^{(k)}} & {{{\bf{I}}_N}} & {{{\bf{0}}_{N \times \left( {L_{mn}^{(\max)} - L_{mn}^{(k)}} \right)}}}  \\
\end{array}} \right] \in {{\mathbb{C}}^{N \times \left( {N + L_{mn}^{(\max)}} \right)}}$. The Nyquist samples from each of the ${N + L_{mn}^{(\max)}}$ 
range-cells for the $q$-th pulse are collected in the following vector
\begin{flalign}
{\bf{y}}_{mn}^{(q)} = {\bf{z}}_{mn}^{\left( q \right)} + {\bf{w}}_{mn}^{\left( q \right)}, \;q=0,\cdots,Q-1, \label{eq_receive_signal_discretize}
\end{flalign}
where ${\bf{w}}_{mn}^{\left( q \right)}$ is the sampled noise vector and the signal trail is 
\begin{align}
{\bf{z}}_{mn}^{\left( q \right)} = 
&\sum\limits_{k = 1}^K {\sqrt E}{\beta _{mn}^{(k)}\left( q \right){e^{\mathrm{j}2\pi f_{mn}^{(k)}q{T_{\rm PRI}}}}{\bf{C}}_{L_{mn}^{(k)}}^T{{\bf{s}}_m}} \nonumber \\
= & {{\bf{A}}_{mn}}{{\bf{x}}_{mn}^{\left( q \right)}},
\end{align}
where ${{\bf{A}}_{mn}} = \left[ {C_{L_{mn}^{\left( 1 \right)}}^T{{\bf{s}}_m}, \cdots ,C_{L_{mn}^{\left( K \right)}}^T{{\bf{s}}_m}} \right] \in {{\mathbb{C}}^{\left( {N + L_{mn}^{\max }} \right) \times K}}$ and ${{\bf{x}}_{mn}^{\left( q \right)}} = {\left[ {\beta _{mn}^{\left( 1 \right)}{e^{\mathrm{j}2\pi f_{mn}^{\left( 1 \right)}q{T_{\rm PRI}}}}, \cdots ,\beta _{mn}^{\left( K \right)}{e^{\mathrm{j}2\pi f_{mn}^{\left( K \right)}q{T_{\rm PRI}}}}} \right]^T}$.
Here, ${{\bf{s}}_m} \in {{\mathbb{C}}^{N \times 1}}$ is the sampled transmit waveform  from the $m$-th transmit antenna. In the above, we used ${e^{\mathrm{j}2\pi f_{mn}^{(k)}t}}\approx{e^{\mathrm{j}2\pi f_{mn}^{(k)}q{T_{\rm PRI}}}}$ which follows from the condition \textbf{C3}. 

After collecting samples for $Q$ pulses, we formulate the noise-free signal trail of the 
data matrix at the $n$-th receiver as 
\begin{align}
{{\bf{Z}}_{mn}} = {\left[ {{\bf{z}}_{mn}^{(1)},{\bf{z}}_{mn}^{(2)}, \cdots ,{\bf{z}}_{mn}^{(Q)}} \right]^T} =   {{\bf{D}}_{mn}}{{\bf{\Lambda }}_{mn}}{{\bf{\Gamma }}_{mn}}, \label{eq_Z_data_matrix}
\end{align}
where the transmit signal matrix 
${{\bf{\Gamma }}_{mn}} = {\left[ {\begin{array}{*{20}{c}}
   {{\bf{C}}_{L_{mn}^{(1)}}^T{{\bf{s}}_m}} &  \cdots  & {{\bf{C}}_{L_{mn}^{(K)}}^T{{\bf{s}}_m}}  \\
\end{array}} \right]^T} $,  the Doppler matrix ${{\bf{D}}_{mn}} = \left[ {{\bf{d}}_{mn}^{\left( 1 \right)}, \cdots ,{\bf{d}}_{mn}^{\left( K \right)}} \right]$ and the reflectivity matrix ${{\bf{\Lambda }}_{mn}} = {\rm{diag}}\left\{ {\left[ {\beta _{mn}^{\left( 1 \right)}, \cdots ,\beta _{mn}^{\left( K \right)}} \right]} \right\}$.
Here, $ {\bf{d}}_{mn}^{(k)}$ 
is the Doppler steering vector defined as \cite{he2010mimo}
\begin{align}
 {\bf{d}}_{mn}^{(k)} = {\left[ {1,{e^{\mathrm{j}2\pi f_{mn}^{(k)}{T_{\rm PRI}}}}, \cdots ,{e^{\mathrm{j}2\pi f_{mn}^{(k)}\left( {Q - 1} \right){T_{\rm PRI}}}}} \right]^T}.
\end{align}
We have the following result regarding the rank of the noise-free data matrix ${\bf {Z}}_{mn}$.

\begin{proposition}
\label{prop:low_rank}
and ${\bf {Z}}_{mn}  \in {{\mathbb{C}}^{Q \times \left( {N + L_{mn}^{(\max)}} \right)}}$ be the data matrix formulated from the samples at the $n$-th receive antenna for the reflected echoes corresponding to the $m$-th transmit signal. Then, the rank of ${\bf {Z}}_{mn}$ is determined by the number of different ranges as well as different velocities among all targets and the rank is bounded by $K$. 
\end{proposition}
\begin{IEEEproof}
The matrices ${{\bf{\Gamma }}_{mn}}$ and ${{\bf{D}}_{mn}}$ in (\ref{eq_Z_data_matrix}) have the dimensions ${K \times \left( {N + L_{mn}^{(\max)}} \right)}$ and $Q\times K$, respectively. Under the assumption of slow moving targets, each target stays in the same range bin during a CPI.
We note that the rank of matrix ${{\bf{D}}_{mn}}$  is governed by the velocity differences among all $K$ targets. If more than one targets have the same velocity, the rank of matrix ${{\bf{D}}_{mn}}$ could be less than $K$. For matrix ${{\bf{\Gamma }}_{mn}}$, its rank is governed by the time-differences-of-arrival   $L_{mn}^{(K)}, k=1,\cdots,K$, which, in turn, are determined by the range differences of all $K$ targets. If more than one targets occupy the same range bin, the rank of ${{\bf{\Gamma }}_{mn}}$ could be less than $K$. Therefore, following (\ref{eq_Z_data_matrix}), the data matrix 
${{\bf{Z}}_{mn}} \in {{\mathbb{C}}^{Q \times \left( {N + L_{mn}^{(\max)}} \right)}}$ is low rank and its rank is bounded by $K$.
\end{IEEEproof}

Combining the noise-trail with ${{\bf{Z}}_{mn}}$, we obtain the full data matrix
\begin{align}
{{\bf{Y}}_{mn}} = {{\bf{Z}}_{mn}} + {{\bf{W}}_{mn}}, 
\end{align}
where ${{\bf{W}}_{mn}}$ is the sampled noise matrix.

\subsection{Reduced-Rate Sampling and Matrix Completion}
In our MC-WS-MIMO, each receiver samples the incoming signal during each pulse at sub-Nyquist rates. There are several ways to implement a sub-Nyquist sampler \cite{mishra2019sub}. Here, we assume that the samples are selected uniformly at random. For each Tx-Rx pair, these low-rate samples are modeled as partially observed data matrices ${{\bf{X}}_{mn}} \in {{\mathbb{C}}^{Q \times \left( {N + L_{mn}^{(\max)}} \right)}}$:
\begin{align}
{\left[ {\bf{X}}_{mn} \right]_{ij}} = \left\{ \begin{array}{l}
 {\left[ {\bf{Y}}_{mn} \right]_{ij}},\quad\;\left( {i,j} \right) \in \Omega,  \\ 
 0,\qquad\qquad\; {\rm{otherwise},} \\ 
 \end{array} \right.
\end{align}
where $\Omega$ is the set of indices of observed entries with $|\Omega|=h$. The above sampling process can be compactly represented by using the operator ${{\mathcal{P}}_\Omega }$ such that ${\bf{X}}_{mn} = {{\mathcal{P}}_\Omega }\left( {\bf{Y}}_{mn} \right)$.

The receiver then forwards these partially observed data matrices to a fusion center which recovers the missing entries by applying MC techniques as follows. 
From Proposition~\ref{prop:low_rank}, each of the matrices ${{\bf{Z}}_{mn}} $, $n=1,\cdots,M_r$, $m=1,\cdots,M_t$, is low rank; their rank being bounded by $K$. 
In the noise-free case, these matrices can be completed by solving the following optimization \cite{candes2009exact}\cite{candes2010power}
 \begin{align}
  &\textrm{minimize} \;\; {\left\| {\mathbf{X}_{mn}} \right\|_*}  \nonumber \\
  &{\textrm{subject to}} \;\; {{\mathcal{P}}_\Omega }\left( {\bf{X}}_{mn} \right) = {{\mathcal{P}}_\Omega }\left( {\bf{Z}}_{mn} \right),  \label{exact_MC_opt}
\end{align}
where the nuclear norm ${\left\|  {\mathbf X}_{mn}  \right\|_*}$ is the sum of singular values of matrix ${\bf X}_{mn}$.

Here, the conditions of MC are related to the bounds on the coherence of ${\bf Z}_{mn}$. Assume the compact singular value decomposition (SVD) of ${\bf Z}_{mn}$ is ${{\bf Z}_{mn}} = \sum\nolimits_{k = 1}^K {{\rho _k}{{\bf{u}}_k}{\bf{v}}_k^H} $, where $\rho_k$, $k=1,\cdots,K$ are the singular values, and ${\bf u}_k$ (${\bf v}_k$) are the corresponding left (right) singular vectors. The subspaces spanned by ${\bf u}_k$ and ${\bf v}_k$ are $U$ and $V$, respectively. Denote $n_1=Q$ and $n_2=N + L_{mn}^{(\max)}$. The coherence of $U$ (and similarly for $V$) is \cite{candes2009exact}    
\begin{align}
\mu \left( U \right) = \frac{{{n_1}}}{K}\mathop {\max }\limits_{1 \le i \le {n_1}} {\left\| {{{\bf{U}}^{\left( i \right)}}} \right\|^2} \in \left[ {1,\frac{{{n_1}}}{K}} \right],
\end{align}
where ${{{\bf{U}}^{\left( i \right)}}}$ is the $i$-th row of matrix ${\bf{U}} = \left[ {{{\bf{u}}_1}, \cdots ,{{\bf{u}}_K}} \right]$. The matrix ${\bf Z}_{mn}$ has coherence with parameters $\mu_0$ and $\mu_1$ if\\
\textbf{A0} $\max \left( {\mu \left( U \right),\mu \left( V \right)} \right) \le {\mu _0}$ for some positive $\mu_0$.\\
\textbf{A1} The maximum element of matrix $\sum\nolimits_{1 \le i \le K} {{{\bf{u}}_i}{\bf{v}}_i^H} $ is bounded by ${\mu _1}\sqrt {{K \mathord{\left/
 {\vphantom {K {\left( {{n_1}{n_2}} \right)}}} \right.
 \kern-\nulldelimiterspace} {\left( {{n_1}{n_2}} \right)}}}$ in absolute value for some positive $\mu_1$.

If the matrix ${\bf Z}_{mn}$ satisfies \textbf{A0} and \textbf{A1}, the following theorem provides a probabilistic bound for the number of observed entries needed to successfully recover matrix ${\bf Z}_{mn}$.
\begin{theorem}
\cite{candes2009exact} Suppose we observe $h$ entries of a rank-$K$ matrix of ${\bf{Z}}_{mn} \in {{\mathbb{C}}^{{n_1} \times {n_2}}}$ uniformly at random. Assume $b = \max \left( {{n_1},{n_2}} \right)$. There exist constants $C$ and $\zeta$ that if 
\begin{align}
h \ge C\max \left( {\mu _1^2,\mu _0^{{1 \mathord{\left/
 {\vphantom {1 2}} \right.
 \kern-\nulldelimiterspace} 2}}{\mu _1},{\mu _0}{b^{{1 \mathord{\left/
 {\vphantom {1 4}} \right.
 \kern-\nulldelimiterspace} 4}}}} \right)\gamma K b\log b,
\end{align}
for some $\gamma > 2$, the minimizer to problem (\ref{exact_MC_opt}) is unique  and equal to ${\bf Z}_{mn}$ with probability of $1 - \zeta {b^{ - \gamma }}$. For $K \le \mu _0^{ - 1}{b^{{1 \mathord{\left/
 {\vphantom {1 5}} \right.
 \kern-\nulldelimiterspace} 5}}}$, the bound can be improved to $m \ge C{\mu _0}{b^{{6 \mathord{\left/
 {\vphantom {6 5}} \right.
 \kern-\nulldelimiterspace} 5}}}\gamma K \log b$, without affecting the probability of success.
\end{theorem}

In the presence of noise, we have ${{\mathcal{P}}_\Omega }\left( {\bf{Y}}_{mn} \right) = {{\mathcal{P}}_\Omega }\left( {\bf{Z}}_{mn} \right) + {{\mathcal{P}}_\Omega }\left( {\bf{W}}_{mn} \right)$. Then, ${\bf Z}_{mn}$ is completed by solving the optimization
 \begin{align}
  &\textrm{minimize} \;\;{\left\| {\mathbf{X}_{mn}} \right\|_*}  \nonumber \\
  &{\textrm{subject to}} \;\; {\left\| {{{\mathcal{P}}_\Omega }\left( {{\bf{X}}_{mn} - {\bf{Y}}_{mn}} \right)} \right\|_{\mathcal{F}}} \le \delta,  \label{MC_opt_with_noise}
\end{align}
 where ${\delta ^2} = \left( {h + \sqrt {8h} } \right){\sigma ^2}$ and $\sigma^2$ is the covariance of noise. Denote he solution to the optimization problem (\ref{MC_opt_with_noise}) by ${{\bf{\hat Z}}_{mn}}$. Then, the error norm is bounded as ${\left\| {{\bf{\hat Z}}_{mn} - {\bf{Z}}_{mn}} \right\|_\mathcal{F}} \le 4\sqrt {{{\left( {2{n_1}{n_2} + m} \right)\min \left( {{n_1},{n_2}} \right)} \mathord{\left/
 {\vphantom {{\left( {2{n_1}{n_2} + m} \right)\min \left( {{n_1},{n_2}} \right)} m}} \right.
 \kern-\nulldelimiterspace} m}} \delta  + 2\delta $ \cite{candes2010matrix}. The common singular value thresholding (SVT) algorithm \cite{cai2010svt} can be applied to solve the above nuclear norm problem.

\section{Target Localization}
\label{sec:tgt_loc}

\color{black}
Once the matrices ${\mathbf Z}_{mn}, m=1,\cdots,M_t, n=1,\cdots,M_r$ are recovered via MC technique at the fusion center, the unknown target parameters are estimated using any of the classical signal processing techniques such as ML \cite{he2010noncoherent}, least squares, or sparse reconstruction methods \cite{gogineni2011target}. Since the target parameters in a WS-MIMO are usually statistically modeled, we adopt the ML approach for target localization here. 

\subsection{Maximum Likelihood Method}
Denote the unknown target parameters by ${\bm{\theta }} = {\left[ {x,y} \right]^T}  \in \Theta $, 
where $\Theta$ is a two-dimensional space that includes all possible values of $\left( {x,y} \right)$. Assume the hypotheses $\mathcal{H}_1$ and $\mathcal{H}_0$ correspond to, respectively, the presence and absence of the target return in the received signal in (\ref{eq_receive_signal_discretize}) that follows the distribution
\begin{align}
{\bf{y}}_{mn}^{\left( q \right)} \sim {{\mathcal{N}}_C}\left( {{\bf{z}}_{mn}^{\left( q \right)},\sigma _n^2{\bf{I}}} \right),
\end{align}
where ${{\mathcal{N}}_C}$ denotes the complex multivariate circularly symmetric Gaussian probability density function.
The negative log-likelihood ratio (LLR) of hypotheses $H_1$ and $H_0$, is 
\begin{align}
{{\mathcal{L}}_{mn}}\left( {{\bm{\theta }},{{\bf{Y}}_{mn}}} \right) = \frac{1}{{\sigma _n^2}}\sum\limits_{q = 0}^{Q-1} {{{\left\| {{\bf{y}}_{mn}^{\left( q \right)} - {{\bf{A}}_{mn}}{{\bf{x}}_{mn}^{\left( q \right)}}} \right\|}^2}} . \label{eq_LRR}
\end{align}
Since the noise and target reflection coefficients are statistically independent, the joint likelihood ratio is the product of individual likelihood ratios. The joint negative LLR is 
\begin{align}
{\mathcal{L}}\left( {\bm{\theta }} \right) &=  \sum\limits_{m = 1}^{{M_t}} {\sum\limits_{n = 1}^{{M_r}} {{{\mathcal{L}}_{mn}}\left( {{\bm{\theta }},{{\bf{Y}}_{mn}}} \right)} } \nonumber \\
&= \sum\limits_{m = 1}^{{M_t}} {\sum\limits_{n = 1}^{{M_r}} {\sum\limits_{q = 0}^{Q - 1} \frac{1}{{\sigma _n^2}} {{{\left\| {{\bf{y}}_{mn}^{\left( q \right)} - {{\bf{A}}_{mn}}{\bf{x}}_{mn}^{\left( q \right)}} \right\|}^2}} } } .\label{eq_joint_LRR}
\end{align}
By minimizing (\ref{eq_joint_LRR}) over ${{\bf{x}}_{mn}^{\left( q \right)}}$, the least squares solution is
\begin{align}
{{{\bf{\hat x}}}_{mn}^{\left( q \right)}} = {{\bf{A}}_{mn}}{\left( {{\bf{A}}_{mn}^H{{\bf{A}}_{mn}}} \right)^{ - 1}}{\bf{A}}_{mn}^H{\bf{y}}_{mn}^{\left( q \right)}. \label{eq_LS_xq}
\end{align}
By substituting (\ref{eq_LS_xq}) in (\ref{eq_joint_LRR}), the joint negative LLR function becomes
\begin{align}
{\mathcal{L}}\left( {\bm{\theta }} \right) = \sum\limits_{m = 1}^{{M_t}} {\sum\limits_{n = 1}^{{M_r}} {\sum\limits_{q = 0}^{Q - 1} \frac{1}{{\sigma _n^2}} {{{\left\| {{\bf{P}}_{mn}^ \bot {\bf{y}}_{mn}^{\left( q \right)}} \right\|}^2}} } }  ,
\end{align}
where ${\bf{P}}_{mn}^ \bot  = {\bf{I}} - {{\bf{A}}_{mn}}{\left( {{\bf{A}}_{mn}^H{{\bf{A}}_{mn}}} \right)^{ - 1}}{\bf{A}}_{mn}^H$ is the orthogonal projection matrix on the column space of ${{\bf{A}}_{mn}}$.
The ML estimate of the parameter vector $\bm {\theta}$ is 
\begin{align}
{{{\bm{\hat \theta }}}_{\rm ML}} = \arg \mathop {\max }\limits_\Theta  \left( { - \sum\limits_{m = 1}^{{M_t}} {\sum\limits_{n = 1}^{{M_r}} {\sum\limits_{q = 0}^{Q - 1} \frac{1}{{\sigma _n^2}} {{{\left\| {{\bf{P}}_{mn}^ \bot {\bf{y}}_{mn}^{\left( q \right)}} \right\|}^2}} } } } \right). \label{eq_arg_max_LLR} 
\end{align}
In general, the computationally demanding  problem in (\ref{eq_arg_max_LLR}) is solved by nonlinear optimization algorithms such as genetic algorithms and simulated annealing \cite{hassanien2012moving}. In this paper, we adopt a two-dimensional search over $\Theta$ to find the peaks of $-{\mathcal{L}}\left( {\bm{\theta }} \right)$.

\subsection{Geometric Method}

Alternatively, a geometric approach may be employed for localization \cite{noroozi2015target} to obtain a closed-form solution that leads to a reduced computational complexity when compared with the ML approach. For example, \cite{noroozi2015target} employs a two-stage weighted least squares (WLS) to determine the location of a target based on the bistatic range measurements in a passive MIMO radar. Similar closed-form localization algorithms have been suggested for distributed MIMO radars, wherein the transmitters and receivers may or may not be co-located \cite{park2015closed}. These approaches require time delay (TD) estimation to get an initial measurement of the range from transmitters to receivers. 

Assume that the unknown target position is ${\bf{p}}^{(m)}= \left[ {x_{t}^{(m)},y_{t}^{(m)}} \right]^T$, the $i$-th transmitter placed at known position ${\bf{p}}_t^{(i)}= \left [ x_{T_{i}} ,y_{T_{i}}  \right ] ^{\mathrm{T}}$ for $ i = 1,2,...,M_{t}$, and the $j$-th receiver at known positions ${\bf{p}}_r^{(j)}= \left [ x_{R_{j}} ,y_{R_{j}}  \right ] ^{\mathrm{T}}$ for $ j = 1,2,...,M_{r}$. The distance between the target and the $i$-th transmitter is
\begin{align}
R_{tT_{i}} =  \sqrt{\left ( x_{T_{i}}-x_{t} \right )^{2}+\left ( y_{T_{i}}-y_{t} \right )^{2} }.
\end{align}
The distance between the target and the $j$-th receiver is
\begin{align}
R_{tR_{j}} =  \sqrt{\left ( x_{R_{j}}-x_{t} \right )^{2}+\left ( y_{R_{j}}-y_{t} \right )^{2} }.
\end{align}
Then, the total range $ R_{tT_{i}R_{j}} $ is
\begin{align}
 R_{tT_{i}R_{j}} &=  R_{tT_{i}} + R_{tR_{j}} 
 \nonumber \\
 & = \sqrt{\left ( x_{T_{i}}-x_{t} \right )^{2}+\left ( y_{T_{i}}-y_{t} \right )^{2} } \\
 & + \sqrt{\left ( x_{R_{j}}-x_{t} \right )^{2}+\left ( y_{R_{j}}-y_{t} \right )^{2} }.
\label{eq_br_range} 
\end{align} 
Reformulate (\ref{eq_br_range}) as
\begin{align}
& R_{tT_{i}R_{j}} - \sqrt{\left ( x_{T_{i}}-x_{t} \right )^{2}+\left ( y_{T_{i}}-y_{t} \right )^{2} } 
\nonumber \\
&= \sqrt{\left ( x_{R_{j}}-x_{t} \right )^{2}+\left ( y_{R_{j}}-y_{t} \right )^{2}}.
\label{eq_br_range_2} 
\end{align} 
Squaring both sides of (\ref{eq_br_range_2}), rearranging the terms, and simplifying yields
\begin{align}
& \left ( x_{R_{j}}-x_{T_{i}}\right )x_{t} + \left ( y_{R_{j}}-y_{T_{i}} \right )y_{t} -  R_{tT_{i}R_{j}}R_{tT_{i}}
\nonumber \\
&= \frac{1}{2} \left ( R_{R_{j}}^{2} -R_{tT_{i}R_{j}}^{2} - R_{T_{i}}^{2} \right ),
\label{eq_br_range_3} 
\end{align}
where $R_{R_{j}}$ ($R_{tT_{i}R_{j}}$) are the position coordinates of transmitter (receiver):
\begin{align}
R_{T_{i}} &=\sqrt{ x_{T_{i}}^{2} + y_{T_{i}}^{2}},
\end{align}
and
\begin{align}
R_{R_{j}} &=\sqrt{ x_{R_{j}}^{2} + y_{R_{j}}^{2}}, 
\end{align}
Arranging equation (\ref{eq_br_range_3}) in a matrix form with $ M_{t}$ transmitters and $ M_{r}$ receivers leads to the following system of linear equations:
\begin{align}
\mathbf{Ax} =\mathbf{b},
\label{eq_br_range_matrix_form} 
\end{align}
where
\begin{align}
\mathbf{A} &=\begin{bmatrix}
  \mathbf{D}_{1}&  -\mathbf{r} _{t1} &  \dots& \mathbf{0}  \\
  \mathbf{D}_{2}&  \mathbf{0} & \dots & \mathbf{0}  \\
  \vdots &  \vdots & \ddots  & \vdots  \\
  \mathbf{D}_{M_{t}}&  \mathbf{0}&  \dots & -\mathbf{r} _{tM_{t}} 
\end{bmatrix} \in \mathbb{C}^{\left ( M_{t}\times M_{r} \right )\times \left ( M_{t}+ 3 \right )}, \\ 
\mathbf{x} &= \begin{bmatrix}
 x_{t} & y_{t}  & R_{tT_{1}} & \cdots  & R_{tT_{M_{t}}} 
\end{bmatrix}^{\mathrm{T} }, \\ 
\mathbf{b} & = \begin{bmatrix}
 \mathbf{b}_{1}^{\mathrm{T}} & \mathbf{b}_{2}^{\mathrm{T}} & \mathbf{b}_{3}^{\mathrm{T}} & \cdots  & \mathbf{b}_{M_{t}}^{\mathrm{T}} \end{bmatrix}^{\mathrm{T} }, \\ \mathbf{D}_{i} &=\begin{bmatrix}
  {x}_{R_{1}}-{x}_{T_{i}}&  {y}_{R_{1}}-{y}_{T_{i}}  \\
   {x}_{R_{2}}-{x}_{T_{i}}&  {y}_{R_{2}}-{y}_{T_{i}}  \\
  \vdots &  \vdots  \\
  {x}_{R_{M_{r}}}-{x}_{T_{i}}&  {y}_{R_{M_{r}}}-{y}_{T_{i}}
  \end{bmatrix} \in \mathbb{C}^{M_{r} \times 2}, \\ 
  \mathbf{b}_{i} &= \frac{1}{2} \begin{bmatrix}
 R_{R_{1}}^{2} -R_{tT_{i}R_{1}}^{2} - R_{T_{i}}^{2} \\
 \vdots  \\
 R_{R_{M_{r}}}^{2} -R_{tT_{i}R_{M_{r}}}^{2} - R_{T_{i}}^{2}
\end{bmatrix}\in \mathbb{R}^{M_{r} \times 1},\;i=1, \cdots,M_t,
\end{align}
and $ \mathbf{r} _{ti} = \left [ R_{tT_{i}R_{1}},...,R_{tT_{i}R_{M_{r}}} \right ]^{\mathrm{T}}  $ is the range measurement vector corresponding to the $ i$-th transmit antenna, and $ \mathbf{r} _{t} = \left [ \mathbf{r} _{t1},...,\mathbf{r}_{tM_{t}} \right ]^{\mathrm{T}}$ is the range measurement vector for all transmit antennas. Applying the ML method to TD estimation produces \cite{tajer2010optimal} 
\begin{align}
\hat{\bm{\tau}} =\mathrm{arg}  \underset{\mathbf{\bm{\tau}}}{\mathrm{max} } \frac{\left | \sum\limits_{m=1}^{M_{t}}\sum\limits_{n=1}^{M_{r}}\frac{e^{\mathrm{j}2\pi f_{c}\tau_{mn}}}{\tau_{mn}^{\beta}} \sum\limits_{q=0}^{Q}\mathrm{\mathbf{y } }_{mn}^{\left ( q \right ) }\mathrm{\mathbf{s} }_{mn}^{\dagger \left ( q \right ) } \right | ^{2} } {\sum\limits_{m=1}^{M_{t}}\sum\limits_{n=1}^{M_{r}} \frac{1}{\tau_{mn}^{2\beta}}},
\end{align}
where $\mathrm{\mathbf{y}}_{mn}$ is the signal vector transmitted from the $m$-th transmit antenna and received by the $n$-th receive antenna, $ \mathrm{\mathbf{s}}_{m}^{\dagger }$ is conjugate transposition of transmit waveform vector of $ \mathrm{\mathbf{s}}_{m}$ transmitted from the $m$-th transmit antenna, $\beta$ is path loss. 
After obtaining the TD estimates, applying WLS yields  
\begin{align}
\hat{\mathbf{x}}_{\textrm{init}} =\left ( \hat{\mathbf{A}}^{\mathrm{T} }  \hat{\mathbf{A}} \right ) ^{-1}  \hat{\mathbf{A}}^{\mathrm{T
}}\mathbf{b},
\end{align}
from which we obtain the initial target position $\left [ \hat{x}_{t}, \hat{y}_{t} \right ]^{\mathrm{T} } $ as 
\begin{align}
\begin{bmatrix}
\hat{x}_{t}
 \\
\hat{y}_{t}
\end{bmatrix} = \begin{bmatrix}
  1&  0&  0&  0& \dots & 0\\
  0&  1&  0&  0& \dots & 0
\end{bmatrix}\hat{\mathbf{x}}_{\textrm{init}}.
\end{align}
The distance between the target and transmitters used for designing the weighting matrix is then computed using $\left [ \hat{x}_{t}, \hat{y}_{t} \right ]^{\mathrm{T} } $. The weighting matrix $\mathbf{W}$ becomes
\begin{align}
{\mathbf{W}} = {\bm{\mathrm{C}}_{\bm{\varepsilon}}^{-1}} = \left ( \left(\bm{\Lambda} -\bm{\Gamma}\right ) {\bm{\mathrm{C}}_{\bm{n}}}\left ( \bm{\Lambda} -\bm{\Gamma}\right )\right)^{-1},
\end{align}
where $ \bm{\Lambda} = \mathrm{diag\left ( \mathbf{r} _{t}  \right ) } $, $\bm{\Gamma} = \left ( \left [ R_{tT_{1}}\mathbf{1}^{\mathrm{T}},..., R_{tT_{M_{t}}}\mathbf{1}^{\mathrm{T}}\right ]^{\mathrm{T}}  \right ) $, and $ \mathbf{1} $ is a vector of all ones of length $ M_{r}$. The target location estimate is
\begin{align}
\hat{\mathbf{x}}_{\textrm{final}} =\left ( \hat{\mathbf{A}}^{\mathrm{T} }  {\mathbf{W}}\hat{\mathbf{A}} \right ) ^{-1}  \hat{\mathbf{A}}^{\mathrm{T
}}\mathbf{W}\mathbf{b}.
\end{align}

\subsection{Lower Error Bounds} 
Define the target parameter vector $\bm{\theta} = \left [ \hat{x} _{t},\hat{y}_{t},\hat{\xi}_{re}, \hat{\xi}_{im} \right ]  $, where $\hat{\xi}_{re}$ ($\hat{\xi}_{im}$) is the real (and imaginary) part of target's complex reflectivity, that needs to be estimated. The CRLB  for estimating $\bm{\theta}$ is \cite{kay1993fundamentals}
\begin{align}
\mathbf{CRLB}  = \mathbf{J} ^{-1} \left ( \bm{\theta}   \right ),
\end{align}
where 
the Fisher information matrix (FIM) is
\begin{align}
\mathbf{ J} \left ( \bm{\theta}\right ) =E_{\mathbf{y} \mid \bm{\theta} }\left \{ \left [  \frac{\partial }{\partial \bm{\theta} } \mathrm{log} p\left ( \mathbf{y} \mid \bm{\theta}  \right ) \right ] \left [  \frac{\partial }{\partial \bm{\theta} } \mathrm{log} p\left ( \mathbf{y} \mid \bm{\theta}  \right ) \right ]^{\mathrm{T}}  \right \},
\end{align}
where $ \mathrm{log} p\left ( \mathbf{y} \mid \bm{\theta}  \right ) $ is the probability density function (pdf) of received signal $ \mathbf{y}$ conditioned on $\bm{\theta}$ and $ E_{\mathbf{y} \mid \bm{\theta} }\left \{ \cdot  \right \}$ is the conditional expectation of $ \mathbf{y}$ given $ \bm{\theta} $. We are interested in only the target position and, thus, need to extract the $2 \times 2 $ submatrix of CRLB matrix, i.e., 
$ \left [ \mathbf{CRLB} \right ]_{2 \times 2} = \left[ \mathbf{J} ^{-1} \left ( \bm{\theta}   \right ) \right]_{2 \times 2}$. Using the Schur complement of block matrix \cite{golub2013matrix},
\begin{align}
\left [ \mathbf{CRLB} \right ]_{2 \times 2} = \left ( \mathbf{H} \bm{\Lambda} \mathbf{H} ^{\mathrm{T} }  - \mathbf{VH} \bm{\Lambda}_{\zeta }^{\mathrm{T} } \mathbf{H} ^{\mathrm{T} }\mathbf{V} ^{\mathrm{T} }  \right )^{-1},
\end{align}
where 
\begin{align}
\mathbf{H} = \begin{bmatrix}
  A_{11}& \dots & A_{M_{r}M_{t}}\\
  B_{11}& \dots & B_{M_{r}M_{t}}
\end{bmatrix}_{M_{r}M_{t} \times M_{r}M_{t}},
\end{align}
\begin{align}
\bm{\Lambda} = 8\pi^{2}\mathrm{SNR}\left ( f_{c}^{2}+\beta^{2}\right ) \mathbf{I}_{M_{r}M_{t} \times M_{r}M_{t}},
\end{align}
\begin{align}
\mathbf{H} = \frac{4\pi f_{c}\mathrm{SNR}}{\left | \zeta  \right |^{2}}\begin{bmatrix}
 -\zeta_{im}   & \zeta_{re} \\
\vdots    & \vdots\\
-\zeta_{im}  & \zeta_{re}
\end{bmatrix}_{M_{r}M_{t} \times 2},
\end{align}
and
\begin{align}
\bm{\Lambda}_{\zeta } = \mathrm{SNR}\frac{M_{t}M_{r}}{\left | \zeta  \right |^{2} } \mathbf{I}_{2 \times 2},
\end{align}
where 
$\mathrm{SNR}=\frac{E}{{\left | \zeta  \right |^{2} }}$, and 
$\beta = \frac{\int_{B}f^{2}\left | S\left ( f \right )  \right |^{2}df }{\int_{B}\left | S\left ( f \right )  \right |^{2}df}$ is the effective bandwidth and the integration is over the bandwidth $B$. The elements of matrix $\mathbf{H}$ are defined as $ A_{lk} = \cos \phi_{tk} +\cos \phi_{rl} $ and $ B_{lk} = \sin\phi_{tk} + \sin\phi_{rl}$, where $\phi_{tk} = \tan^{-1}\left ( \frac{y_{t}-y_{tk}}{x_{t}-x_{tk}}  \right )$ and $\phi_{rl} = \tan^{-1}\left ( \frac{y_{t}-y_{rl}}{x_{t}-x_{rl}}  \right )$ are the phases that reveal the geometric relationship between the Tx-Rx locations and the target position. 

The minimum mean square errors (MMSEs) in the estimate of the target's x- and y-coordinates are, respectively,
\begin{align}
\sigma_{x}^{2}\left ( \textrm{MMSE} \right ) =  \frac{c^{2}}{8\pi ^{2}SNR\left ( f_{c}^{2}+\beta^{2} \right ) \cdot  \frac{e_{x}}{u_{CRLB}}  },\label{crlb_x} 
\end{align}
and
\begin{align}
\sigma_{x}^{2}\left ( \textrm{MMSE} \right ) =  \frac{c^{2}}{8\pi ^{2}SNR\left ( f_{c}^{2}+\beta^{2} \right ) \cdot \frac{e_{y}}{u_{CRLB}}  }, \label{crlb_y} 
\end{align}
where the coefficients 
\begin{align}
e_{x} =  \left [ \sum_{l=1}^{M_{r}}\sum_{k=1}^{M_{t}}\left ( B_{lk}^{2}-\frac{\left (\sum_{l=1}^{M_{r}}\sum_{k=1}^{M_{t}}\left ( B_{lk}^{2} \right ) \right )^{2} }{\left ( 1+\frac{\beta^{2} }{f_{c}^{2} }  \right ) M_{r}M_{t} }    \right )  \right ],
\end{align}
and
\begin{align}
e_{y} =  \left [ \sum_{l=1}^{M_{r}}\sum_{k=1}^{M_{t}}\left ( A_{lk}^{2}-\frac{\left ( \sum_{l=1}^{M_{r}}\sum_{k=1}^{M_{t}}\left ( A_{lk}^{2} \right ) \right )^{2} }{\left ( 1+\frac{\beta^{2} }{f_{c}^{2} }  \right ) M_{r}M_{t} }  \right )  \right ].
\end{align}

\subsection{WS-MIMO Doppler Estimation}
To compute the Doppler velocities, we adopt the ML approach \cite{he2010target}. Define the unknown parameter vector of the $k$-th target as $ \bm{\theta}^{\left ( k \right )}   =\left [ v_{x}^{\left ( k \right )} ,v_{y}^{\left ( k \right )},\beta_{mn}^{\left ( k \right )}  \right ] $, then the unknown parameter vector of all targets is denoted as $ \bm{\theta}   =\left [\bm{\theta}^{\left ( 1 \right )} ,\bm{\theta}^{\left ( 2 \right )} ,...,\bm{\theta}^{\left ( K \right )}   \right ] $ The joint pdf of the received signal vector $\mathbf{y}_{mn} = [\mathbf{y}_{mn}^{\left ( 0 \right )},\mathbf{y}_{mn}^{\left ( 2 \right )},...,\mathbf{y}_{mn}^{\left ( Q-1 \right )}] $ is \cite{he2010target}
\begin{align}
& p\left (\mathbf{y}_{mn}; \bm{\theta}\right ) \propto  \mathrm{exp} \left \{ -\sum_{m=1}^{M_{t}} \sum_{n=1}^{M_{r}}\int_{T}^{} \left | \sum_{k=1}^{K} y_{mn}^{\left ( k \right ) } \left ( t \right ) - \nonumber \right.\right. \\ 
&\left.\left. \sum_{k=1}^{K}\sum_{q=0}^{Q-1} \sqrt{E}  \beta_{mn}^{\left ( k \right ) }s_{m}\left ( t-\tau _{mn}-qT_{\mathrm{PRI} }\right )e^{j2\pi f_{mn}^{\left ( k \right ) }\left ( v_{x}^{\left ( k \right ) },v_{y}^{\left ( k \right ) }  \right)t}\right |^{2}dt \right\}.\label{joint_pdf}
\end{align}
Following \cite{kay1993fundamentals}, the pdf in (\ref{joint_pdf}) yields the ML estimation of unknown parameters as
\begin{align}
\widehat{\bm{\theta}}_{ML} & = \underset{\bm{\theta}}{\mathrm{argmax}}\left\{ \mathrm{ln}\; p\left (\mathbf{y}_{mn}; \bm{\theta}\right )\right\}  \nonumber \\
& = \underset{\bm{\theta}}{\mathrm{argmax}}\left\{ -\sum_{m=1}^{M_{t}} \sum_{n=1}^{M_{r}}\int_{T}^{} \left | \sum_{k=1}^{K} y_{mn}^{\left ( k \right ) } \left ( t \right ) - \sum_{k=1}^{K}\sum_{q=0}^{Q-1} \sqrt{E}\beta_{mn}^{\left ( k \right ) } \nonumber \right.\right. \\ 
&\hspace{4mm}\left.\left. s_{m}\left ( t-\tau _{mn}-qT_{\mathrm{PRI} }\right )e^{j2\pi f_{mn}^{\left ( k \right ) }\left ( v_{x}^{\left ( k \right ) },v_{y}^{\left ( k \right ) }  \right)t}\right |^{2}dt\right\}.
\end{align}
To simplify the analysis, the complex reflectivity coefficients $ \beta_{mn}^{\left ( k \right )}$ are assumed to be the same for all paths under the assumption that the scatters are isotropic \cite{he2010target}, i.e., $ \beta_{mn}^{\left ( k \right )} = \beta$. Then, the derivative of the log-likelihood function with respect to $ \beta$ vanishes, i.e.,
\begin{align}
 \frac{\partial}{\partial \beta_{mn}^{\left ( k \right )} }\mathrm{ln}\; p\left (\mathbf{y}_{mn}; \bm{\theta}\right ) = 0.
\end{align}
The ML estimate of $ \beta$ becomes
\begin{align}
\widehat{\beta}_{ML} & = \frac{1}{\sqrt{E}\rho}\sum_{m=1}^{M_{r}}  \sum_{n=1}^{M_{t}} \sum_{q = 0}^{Q-1} \sum_{k = 1}^{K} \int_{T}^{}y_{mn}^{\left ( k \right ) } \left ( t \right )s_{m}^{\dagger}\left ( t-\tau _{mn}-qT_{\mathrm{PRI} }\right ) \nonumber \\
&\hspace{4cm} e^{-j2\pi f_{mn}^{\left ( k \right ) }\left ( v_{x}^{\left ( k \right ) },v_{y}^{\left ( k \right ) }  \right)t} dt.
\end{align}
where 
\begin{align}
\rho & = \sum_{m=1}^{M_{r}}  \sum_{n=1}^{M_{t}} \int_{T}^{} \left | \sum_{q = 0}^{Q-1} \sum_{k = 1}^{K} y_{mn}^{\left ( k \right ) } \left ( t \right )s_{m}^{\dagger}\left ( t-\tau _{mn}-qT_{\mathrm{PRI} }\right ) \nonumber \right. \\
& \hspace{4cm}\left. e^{-j2\pi f_{mn}^{\left ( k \right ) }\left ( v_{x}^{\left ( k \right ) },v_{y}^{\left ( k \right ) }  \right)t}\right |^{2}dt.
\end{align}
Expanding the likelihood function yields
\begin{align}
& \mathrm{ln}\; p\left (\mathbf{y}_{mn}; \bm{\theta}\right ) \nonumber\\
&= -\sum_{m=1}^{M_{r}}  \sum_{n=1}^{M_{t}} \int_{T}^{}y_{mn}^{2}dt \nonumber \\
&+ 2\beta\sum_{m=1}^{M_{r}}\sum_{n=1}^{M_{t}} \sum_{q=0}^{Q-1}\sum_{k=1}^{K} \int_{T}^{}\sqrt{E}y_{mn}\left(t\right) s_{m}^{\dagger}\left ( t-\tau _{mn}-qT_{\mathrm{PRI} }\right )\nonumber \\
&\hspace{3cm} e^{-j2\pi f_{mn}^{\left ( k \right ) }\left ( v_{x}^{\left ( k \right ) },v_{y}^{\left ( k \right ) }  \right)t}dt \nonumber \\
& - \beta^{2}E\sum_{m=1}^{M_{r}}  \sum_{n=1}^{M_{t}}\int_{T}^{}\left|\sum_{q=0}^{Q-1}\sum_{k=1}^{K}s_{m}\left ( t-\tau _{mn}-qT_{\mathrm{PRI} }\right) \right. \nonumber  \\ 
& \hspace{3cm} \left. e^{j2\pi f_{mn}^{\left ( k \right ) }\left ( v_{x}^{\left ( k \right ) },v_{y}^{\left ( k \right ) }  \right)t}\right|^{2}dt.\label{logfunction_expand}
\end{align}
Since the first and the last terms in (\ref{logfunction_expand}) are both negative, maximizing the whole likelihood function is equivalent to maximizing the second term. By substituting $\beta$ with $\widehat{\beta}_{ML}$, it derives
\begin{align}
\widehat{\bm{v}}_{ML} & = \underset{\bm{v}}{\mathrm{argmax}}\left\{ \mathrm{ln}\; p\left (\mathbf{y}_{mn}; \bm{v}\right )\right\} \nonumber \\
& \left. s_{m}^{\dagger}\left ( t-\tau _{mn}-qT_{\mathrm{PRI} }\right )e^{-j2\pi f_{mn}^{\left ( k \right ) }\left ( v_{x}^{\left ( k \right ) },v_{y}^{\left ( k \right ) }  \right)t}dt \right\} \nonumber \\
& = \underset{{v_{x}^{\left(k\right)}, v_{y}^{\left(k\right)}}}{\mathrm{argmax}}\left\{2\widehat{\beta}_{ML}\sum_{m=1}^{M_{r}}\sum_{n=1}^{M_{t}} \sum_{q=0}^{Q-1}\sum_{k=1}^{K}\int_{T}^{}\sqrt{E}y_{mn}\left(t\right) \nonumber \right.\\
& \left. s_{m}^{\dagger}\left ( t-\tau _{mn}-qT_{\mathrm{PRI} }\right )e^{-j2\pi f_{mn}^{\left ( k \right ) }\left ( v_{x}^{\left ( k \right ) },v_{y}^{\left ( k \right ) }  \right)t}dt \right\} \nonumber \\
& =  \underset{{v_{x}^{\left(k\right)}, v_{y}^{\left(k\right)}}}{\mathrm{argmax}}\left\{\frac{1}{\rho}\left|\sum_{m=1}^{M_{r}}\sum_{n=1}^{M_{t}} \sum_{q=0}^{Q-1}\sum_{k=1}^{K}\int_{T}^{}y_{mn}\left(t\right) \nonumber \right.\right.\\
& \left.\left. s_{m}^{\dagger}\left ( t-\tau _{mn}-qT_{\mathrm{PRI} }\right )e^{-j2\pi f_{mn}^{\left ( k \right ) }\left ( v_{x}^{\left ( k \right ) },v_{y}^{\left ( k \right ) }  \right)t}dt \right|^{2} \right\}.\label{velocity_ML_criterion}
\end{align}
The discrete form of (\ref{velocity_ML_criterion}) is 
\begin{align}
\widehat{\bm{v}}_{ML} & = \underset{{v_{x}^{\left(k\right)}, v_{y}^{\left(k\right)}}}{\mathrm{argmax}}\left\{\frac{1}{\rho}\left|\sum_{m=1}^{M_{r}}\sum_{n=1}^{M_{t}} \sum_{q=0}^{Q-1}\sum_{k=1}^{K}\mathbf{y}_{mn}(k) \nonumber \right.\right.\\
& \left.\left. \mathbf{s}_{m}^{\dagger}\left(k, q\right)e^{-j2\pi f_{mn}^{\left ( k \right ) }\left ( v_{x}^{\left ( k \right ) },v_{y}^{\left ( k \right ) }  \right)qT_{PRI}} \right|^{2} \right\}. \label{eq_velocity_est_ML}
\end{align}
The two-dimensional search is adopted to obtain the ML estimates of the target velocities.

Algorithm \ref{alg_estimation} below summarizes the estimation procedure of target parameters in MC-WS-MIMO radar. 

\begin{algorithm}[H]
\caption{MC-WS-MIMO radar target parameter estimation}
	\label{alg_estimation}
	\begin{algorithmic}[1]
		\Statex \textbf{Input:} Partially observed samples ${{\bf{X}}_{mn}}$ and  set of indices of observed entries $\Omega$.
		\Statex \textbf{Output:}  Target location $(x,y)$ and velocity $(v_x,v_y)$.

    \State  Recover the full data matrix ${\bf{\hat Z}}_{mn}$ by solving the optimization problem of (\ref{MC_opt_with_noise}) using SVT algorithm.

    \State Estimate the target's location, $(x,y)$, with maximum likelihood approach following equation (\ref{eq_arg_max_LLR}).
    \State Estimate the target's velocity, $(v_x,v_y)$, with maximum likelihood approach following equation (\ref{eq_velocity_est_ML}).
	\end{algorithmic}
\end{algorithm}

\section{Performance Analyses}
\label{sec:perf}
To characterize the performance of MC-WS-MIMO radar, we derive the guarantees on the coherence and recoverability of the data matrix, statistical AF, and lower error bounds on parameter estimates.

\subsection{Coherence and Recoverability of ${{\bf{Z}}_{mn}}$}
\label{sec:coherence}

Recall the following useful result from \cite{wolkowicz1980bounds}:
\begin{theorem} \label{theorem_bound_eigenvalue}
\cite{wolkowicz1980bounds} Assume ${\bf{M}} \in {{\mathbb{C}}^{N \times N}}$ be a matrix with real eigenvalues. Define 
\begin{align}
\tau  \triangleq \frac{{{\rm{tr}}\left( {\bf{M}} \right)}}{N}, \quad \kappa^2  \triangleq \frac{{{\rm{tr}}\left( {{{\bf{M}}^2}} \right)}}{N} - {\tau ^2}.
\end{align}
Then, it holds that
\begin{align}
\tau  - \kappa \sqrt {N - 1}  \le {\lambda _{\min }}\left( {\bf{M}} \right) \le \tau  - \frac{\kappa }{{\sqrt {N - 1} }},  \label{eq_eigenvalue_min} \\
\tau  + \frac{\kappa }{{\sqrt {N - 1} }} \le {\lambda _{\max }}\left( {\bf{M}} \right) \le \tau  + \kappa \sqrt {N - 1}, \label{eq_eigenvalue_max}
\end{align}
where $\lambda _{\min}(\cdot)$ ($\lambda _{\max}(\cdot)$) is the minimum (maximum) eigenvalue of its matrix argument. Further, equality holds on the left (right) of (\ref{eq_eigenvalue_min}) if and only if equality holds on the left (right) of (\ref{eq_eigenvalue_max}) if and only if the $N-1$ largest (smallest) eigenvalues are equal. \label{theorem_eigenvalue}
\end{theorem}

We now state our main performance guarantee for MC-WS-MIMO in the following Theorem~\ref{Coherence_of_data_matrix_theorem}.
\begin{theorem} \label{Coherence_of_data_matrix_theorem}
(Coherence of matrix ${{\bf{Z}}_{mn}}$): Consider the widely separated MIMO radar system as presented in Section \ref{sec:sysmod} and assume the set of target Doppler frequency ${\left\{ {f_{mn}^{\left( k \right)}} \right\}_{k \in {\mathbb{N}}_K^ + }}$ consists of almost surely distinct members.
Define 
\begin{align}
   {\beta _Q}\left( \xi _t \right)  &\triangleq  \mathop {\sup }\limits_{x \in \left[ {{\xi _t},{\frac{1}{2}}} \right]} \frac{{{{\sin }^2}\left( {\pi Qx} \right)}}{{{{\sin }^2}\left( {\pi x} \right)}}, \\
   {\xi _t} &\triangleq \mathop {\min }\limits_{\left( {i,j} \right) \in {\bf{N}}_K^ +  \times {\bf{N}}_K^ + ,i \ne j} g\left( {{T_{{\rm{PRI}}}}\left| {f_{mn}^{\left( i \right)} - f_{mn}^{\left( j \right)}} \right|} \right),
\end{align}
and 
\begin{align}
g\left( x \right) \triangleq \left\{ \begin{gathered}
  \left\lceil x \right\rceil  - x,\left\lceil x \right\rceil  - x \leq \tfrac{1}{2} \hfill \\
  x - \left\lfloor x \right\rfloor , {\rm otherwise}. \hfill \\ 
\end{gathered}  \right.
\end{align}
Consider that the transmit waveforms are unimodular following assumption {\bf C7} and the waveform autocorrelation is denoted as
\begin{align}
    \gamma \left( {{l_{ij}}} \right) = \sum\limits_{k = {l_{ij}} + 1}^N {{s_m}\left( k \right)s_m^*\left( {k - {l_{ij}}} \right)}, 
\end{align}
where ${l_{ij}} = L_{mn}^{\left( i \right)} - L_{mn}^{\left( j \right)}$.
If $ K \le \frac{Q}{{\sqrt {{\beta _Q}\left( {{\xi _t}} \right)} }}$, the coherence of matrix  ${{\bf{Z}}_{mn}}$ satisfies 
\begin{align}
 \mu \left( U \right) &\le \frac{Q}{{Q - \left( {K - 1} \right)\sqrt {{\beta _Q}\left( {{\xi _t}} \right)} }}, \\ 
 \mu \left( V \right) &\le \frac{{N + L_{mn}^{\left( {\max } \right)}}}{{N - \sqrt {K - 1} \sqrt {\sum\limits_{j \ne i}^K {\sum\limits_{i = 1}^K {{{\left| {\gamma \left( {{l_{ij}}} \right)} \right|}^2}} } } }} .
\end{align}
The matrix ${{\bf{Z}}_{mn}}$ obeys the conditions {\bf A0} and {\bf A1} with ${\mu _0} \triangleq \max \left\{ {\mu \left( U \right),\mu \left( V \right)} \right\}$, and ${\mu _1} \triangleq \sqrt K {\mu _0}$ with probability $1$.
\end{theorem}
\begin{IEEEproof}
We prove the bounds on $\mu \left( U \right)$ and $\mu \left( V \right)$ separately as follows.\\
\textit{1) Bound on $\mu \left( U \right)$}: We would like to consider the case where both sets of ranges and velocities consist of distinct members. The compact SVD of ${{\bf{Z}}_{mn}}$ can be written as 
\begin{align}
{{\bf{Z}}_{mn}} = {\bf{U\Sigma }}{{\bf{V}}^H},
\end{align}
where ${\bf{U}} \in {{\mathbb{C}}^{Q \times K}},{\bf{V}} \in {{\mathbb{C}}^{\left( {N + L_{mn}^{\left( {\max } \right)}} \right) \times K}}$ such that ${{\bf{U}}^H}{\bf{U}} = {{\bf{I}}_K},{{\bf{V}}^H}{\bf{V}} = {{\bf{I}}_K}$, and ${\bf{\Sigma }} \in {{\mathbb{R}}^{K \times K}}$ is a diagonal matrix containing the singular values of ${{\bf{Z}}_{mn}}$.
Consider the QR decomposition of ${{\bf{D}}_{mn}}$, i.e., ${{\bf{D}}_{mn}} = {{\bf{Q}}_r}{{\bf{R}}_r}$, where ${{\bf{Q}}_r} \in {{\mathbb{C}}^{Q \times K}}$ is such that ${\bf{Q}}_r^H{{\bf{Q}}_r} \equiv {{\bf{I}}_K}$ and ${{\bf{R}}_r} \in {{\mathbb{C}}^{K \times K}}$ is an upper triangular matrix. Similarly, consider the QR decomposition of ${\bf{\Gamma }}_{mn}^T$, i.e., ${\bf{\Gamma }}_{mn}^T = {{\bf{Q}}_s}{{\bf{R}}_s}$, where ${{\bf{Q}}_s} \in {{\mathbb{C}}^{\left( {N + L_{mn}^{\left( {\max } \right)}} \right) \times K}}$ is such that  ${\bf{Q}}_s^H{{\bf{Q}}_s} \equiv {{\bf{I}}_K}$ and ${{\bf{R}}_s}$ is an upper triangular matrix. The matrix ${{\bf{R}}_r}{{\bf{\Lambda }}_{mn}}{\bf{R}}_s^T \in {{\mathbb{C}}^{K \times K}}$ is rank-$K$ matrix and its SVD can be expressed as ${{\bf{R}}_r}{{\bf{\Lambda }}_{mn}}{\bf{R}}_s^T = {{\bf{Q}}_1}{\bf{\Delta Q}}_2^H$. Here, ${{\bf{Q}}_1} \in {{\mathbb{C}}^{K \times K}}$ is such that ${{\bf{Q}}_1}{\bf{Q}}_1^H = {\bf{Q}}_1^H{{\bf{Q}}_1} = {{\bf{I}}_K}$ (the same holds for ${{\bf{Q}}_2}$) and ${\bf{\Delta }} \in {{\mathbb{R}}^{K \times K}}$ is a non-zero diagonal matrix, containing the singular values of matrix ${{\bf{R}}_r}{{\bf{\Lambda }}_{mn}}{\bf{R}}_s^T$. Thus, it holds that 
\begin{align}
{{\bf{Z}}_{mn}} = {{\bf{Q}}_r}{{\bf{Q}}_1}{\bf{\Delta Q}}_2^H{\bf{Q}}_s^T = {{\bf{Q}}_r}{{\bf{Q}}_1}{\bf{\Delta }}{\left( {{\bf{Q}}_s^*{{\bf{Q}}_2}} \right)^H},
\end{align}
is a valid SVD of ${{\bf{Z}}_{mn}}$ since ${\left( {{{\bf{Q}}_r}{{\bf{Q}}_1}} \right)^H}{{\bf{Q}}_r}{{\bf{Q}}_1} = {{\bf{I}}_K}$ and ${\left( {{\bf{Q}}_s^*{{\bf{Q}}_2}} \right)^H}{\bf{Q}}_s^*{{\bf{Q}}_2} = {{\bf{I}}_K}$. According to the uniqueness of singular values of a matrix, it holds that ${\bf{\Sigma }} = {\bf{\Delta }},{\bf{U}} = {{\bf{Q}}_r}{{\bf{Q}}_1}$ and ${\bf{V}} = {\bf{Q}}_s^*{{\bf{Q}}_2}$.

Denote the $i$-th row of ${{\bf{Q}}_r}$ and ${{\bf{D}}_{mn}}$ as ${\bf{Q}}_r^{\left( i \right)}$ and ${{\bf{D}}_{mn}^{\left( i \right)}}$, respectively. The coherence of the row space of ${{\bf{Z}}_{mn}}$ is 
\begin{align}
 \mu \left( U \right) = &\frac{Q}{K}\mathop {\sup }\limits_{i \in {\mathbb{N}}_Q^ + } \left\| {{\bf{Q}}_r^{\left( i \right)}{{\bf{Q}}_1}} \right\|_2^2 = \frac{Q}{K}\mathop {\sup }\limits_{i \in {\mathbb{N}}_Q^ + } \left\| {{\bf{Q}}_r^{\left( i \right)}} \right\|_2^2 \nonumber\\ 
  = & \frac{Q}{K}\mathop {\sup }\limits_{i \in {\mathbb{N}}_Q^ + } \left\| {{\bf{D}}_{mn}^{\left( i \right)}{\bf{R}}_r^{ - 1}} \right\|_2^2 \nonumber\\ 
  \le & \frac{Q}{K}\mathop {\sup }\limits_{i \in {\mathbb{N}}_Q^ + } \frac{{\left\| {{\bf{D}}_{mn}^{\left( i \right)}} \right\|_2^2}}{{\sigma _{\min }^2\left( {{{\bf{R}}_r}} \right)}},  \nonumber\\
   \le & \frac{Q}{{\sigma _{\min }^2\left( {{{\bf{R}}_r}} \right)}},
\end{align}
where 
\begin{align}
\sigma _{\min }^2\left( {{{\bf{R}}_r}} \right) &= {\lambda _{\min }}\left( {{\bf{R}}_r^H{{\bf{R}}_r}} \right) 
= {\lambda _{\min }}\left( {{\bf{R}}_r^H{\bf{Q}}_r^H{{\bf{Q}}_r}{{\bf{R}}_r}} \right) \nonumber\\ &= {\lambda _{\min }}\left( {{\bf{D}}_{mn}^H{{\bf{D}}_{mn}}} \right).
\end{align}
Here, we use the symbol ${\lambda _{\min }}\left(  \cdot  \right)$ to denote the minimal eigenvalue of a matrix. Thus, 
\begin{align}
\mu \left( U \right) \le \frac{Q}{{{\lambda _{\min }}\left( {{\bf{D}}_{mn}^H{{\bf{D}}_{mn}}} \right)}}. \label{eq_coherence_U_bound}
\end{align}

\textit{2) Bound on $\mu \left( V \right)$}: 
According to (\ref{eq_coherence_U_bound}), we need a strict positive lower bound of ${\lambda _{\min }}\left( {{\bf{D}}_{mn}^H{{\bf{D}}_{mn}}} \right)$, with
\begin{align}
{\bf{D}}_{mn}^H{{\bf{D}}_{mn}} = \left[ {\begin{array}{*{20}{c}}
   Q & {{\delta _{1,2}}} &  \cdots  & {{\delta _{1,K}}}  \\
   {\delta _{1,2}^*} & Q &  \cdots  & {{\delta _{2,K}}}  \\
    \vdots  &  \vdots  &  \ddots  &  \vdots   \\
   {\delta _{1,K}^*} & {\delta _{2,K}^*} &  \cdots  & Q  \\
\end{array}} \right],
\end{align}
where
\begin{align}
{\delta _{i,j}} = \sum\limits_{q = 1}^Q {{e^{\mathrm{j}2\pi q\left( {f_{mn}^{\left( i \right)} - f_{mn}^{\left( j \right)}} \right){T_{{\rm{PRI}}}}}}}, \quad \forall \left( {i,j} \right) \in {{\mathbb{N}}_K} \times {{\mathbb{N}}_K} .
\end{align}
We apply Theorem \ref{theorem_bound_eigenvalue} to matrix ${\bf{M}} \triangleq {\bf{D}}_{mn}^H{{\bf{D}}_{mn}}  \in {{\mathbb{C}}^{N \times N}}$. The trace of ${\bf{M}}$ is $KQ$. Thus, 
\begin{align}
\tau  = \frac{{KQ}}{K} = Q.
\end{align}
Since ${\bf{M}}$ is a Hermitian matrix, it is true that
\begin{align}
 &{\rm{tr}}\left( {{{\bf{M}}^2}} \right) \nonumber \\
 &= \sum\limits_{{k_1} = 1}^K {\sum\limits_{{k_2} = 1}^K {{{\left| {{\delta _{{k_1},{k_2}}}} \right|}^2}} } \nonumber  \\ 
 & = \sum\limits_{{k_1} = 1}^K {\left\{ {{Q^2} + \sum\limits_{ \atopnew{\scriptstyle {k_2} = 1}{ 
  \scriptstyle {k_2} \ne {k_1}} }^K {{{\left| {\sum\limits_{q = 1}^Q {{e^{\mathrm{j}2\pi q\left( {f_{mn}^{\left( {{k_1}} \right)} - f_{mn}^{\left( {{k_2}} \right)}} \right){T_{{\rm{PRI}}}}}}} } \right|}^2}} } \right\}} \nonumber \\ 
  & = \sum\limits_{{k_1} = 1}^K {\left\{ {{Q^2} + \sum\limits_{ \atopnew{\scriptstyle {k_2} = 1}{ 
  \scriptstyle {k_2} \ne {k_1}} }^K {\frac{{{{\sin }^2}\left( {\pi Q\left( {f_{mn}^{\left( {{k_1}} \right)} - f_{mn}^{\left( {{k_2}} \right)}} \right){T_{{\rm{PRI}}}}} \right)}}{{{{\sin }^2}\left( {\pi \left( {f_{mn}^{\left( {{k_1}} \right)} - f_{mn}^{\left( {{k_2}} \right)}} \right){T_{{\rm{PRI}}}}} \right)}}} } \right\}} \nonumber \\ 
  & \triangleq \sum\limits_{{k_1} = 1}^K {\left\{ {{Q^2} + \sum\limits_{ \atopnew{\scriptstyle {k_2} = 1}{ 
  \scriptstyle {k_2} \ne {k_1}} }^K {\phi _Q^2\left( {\left( {f_{mn}^{\left( {{k_1}} \right)} - f_{mn}^{\left( {{k_2}} \right)}} \right){T_{{\rm{PRI}}}}} \right)} } \right\}}, \label{eq_trace_M_square}
\end{align}
where \begin{align}
    {\phi _Q}\left( x \right) \triangleq \frac{{\sin \left( {\pi Qx} \right)}}{{\sin \left( {\pi x} \right)}}, \;  x \in {\mathbb R}, \; Q \in {\mathbb N^+}.
\end{align}
For $x \in \left[ {k,k + \tfrac{1}{2}} \right],\forall k \in {\mathbb{Z}}$, the sequence of $\phi _Q^2\left( x \right)$ is strictly decreasing. Define ${\xi _t} \triangleq \mathop {\min }\limits_{i \ne j} g\left( {\left| {\alpha _i^t - \alpha _j^t} \right|} \right) \in \left[ {0,\tfrac{1}{2}} \right]$, where 
\begin{align}
g\left( x \right) \triangleq \left\{ \begin{gathered}
  \left\lceil x \right\rceil  - x,\left\lceil x \right\rceil  - x \leq \tfrac{1}{2} \hfill \\
  x - \left\lfloor x \right\rfloor , {\rm otherwise}. \hfill \\ 
\end{gathered}  \right.
\end{align}
The upper bound of (\ref{eq_trace_M_square}) is 
\begin{align}
  {\text{tr}}\left( {{{\bf M}^2}} \right) &= \sum\limits_{{k_1} = 1}^K {\left\{ {{Q^2} + \left( {K - 1} \right)\mathop {\sup }\limits_{x \in \left[ {{\xi _t},\tfrac{1}{2}} \right]} \phi _Q^2\left( x \right)} \right\}}  \hfill \nonumber\\
   &\triangleq K{Q^2} + K\left( {K - 1} \right){\beta _{{Q}}}\left( \xi_t \right) .
\end{align}
According to Theorem \ref{theorem_eigenvalue}, 
\begin{align}
    {\lambda _{\min }}\left( {\mathbf{M}} \right) = {\lambda _{\min }}\left( {{\mathbf{D}}_{mn}^H{{\mathbf{D}}_{mn}}} \right) \geq Q - \left( {K - 1} \right)\sqrt {{\beta _{{Q}}}\left( \xi_t \right)} .
\end{align}
Therefore, if $K \leq \frac{Q}{{\sqrt {{\beta _{{Q}}}\left( \xi_t \right)} }}$, it holds that 
\begin{align}
    \mu \left( U \right) \leq \frac{Q}{{Q - \left( {K - 1} \right)\sqrt {{\beta _{{Q}}}\left( \xi _t \right)} }}.
\end{align}

The coherence of the column space of ${{\bf{Z}}_{mn}}$  is
\begin{align}
 \mu \left( V \right) & = \frac{{N + L_{mn}^{\left( {\max } \right)}}}{K}\mathop {\sup }\limits_{i \in {\mathbb{N}}_{N + L_{mn}^{\left( {\max } \right)}}^ + } \left\| {{{\bf{Q}}_s^{*\left( i \right)}}{{\bf{Q}}_2}} \right\|_2^2  \nonumber\\ 
 & = \frac{{N + L_{mn}^{\left( {\max } \right)}}}{K}\mathop {\sup }\limits_{i \in {\mathbb{N}}_{N + L_{mn}^{\left( {\max } \right)}}^ + } \left\| {{{\bf{Q}}_s^{\left( i \right)}}} \right\|_2^2 \nonumber \\ 
 & = \frac{{N + L_{mn}^{\left( {\max } \right)}}}{K}\mathop {\sup }\limits_{i \in {\mathbb{N}}_{N + L_{mn}^{\left( {\max } \right)}}^ + } \left\| {{{\left( {{\bf{\Gamma }}_{mn}^T} \right)}^{\left( i \right)}}{\bf{R}}_s^{ - 1}} \right\|_2^2 \nonumber\\ 
 & \le \frac{{N + L_{mn}^{\left( {\max } \right)}}}{K}\frac{{\mathop {\sup }\limits_{i \in {\mathbb{N}}_{N + L_{mn}^{\left( {\max } \right)}}^ + } \left\| {{{\left( {{\bf{\Gamma }}_{mn}^T} \right)}^{\left( i \right)}}} \right\|_2^2}}{{\sigma _{\min }^2\left( {{{\bf{R}}_s}} \right)}}, 
\end{align}
where
\begin{align}
 \sigma _{\min }^2\left( {{{\bf{R}}_s}} \right) & = {\lambda _{\min }}\left( {{\bf{R}}_s^H{{\bf{R}}_s}} \right) = {\lambda _{\min }}\left( {{\bf{R}}_s^H{\bf{Q}}_s^H{{\bf{Q}}_s}{{\bf{R}}_s}} \right) \nonumber \\ 
 & = {\lambda _{\min }}\left( {{{\left( {{\bf{\Gamma }}_{mn}^T} \right)}^H}{\bf{\Gamma }}_{mn}^T} \right). 
\end{align}
Define ${\mathbf{\Phi }} = {\left( {{\mathbf{\Gamma }}_{mn}^T} \right)^H}{\mathbf{\Gamma }}_{mn}^T$. It holds that
\begin{align}
{\mathbf{\Phi }} = \left[ {\begin{array}{*{20}{c}}
   {{\gamma }\left( 0 \right)} & {{\gamma ^*}\left( {{l_{12}}} \right)} &  \cdots  & {{\gamma ^*}\left( {{l_{1K}}} \right)}  \\ 
   {\gamma \left( {{l_{12}}} \right)} & {{\gamma }\left( 0 \right)} &  \cdots  & {{\gamma ^*}\left( {{l_{2K}}} \right)}  \\ 
    \vdots  &  \vdots  &  \ddots  &  \vdots   \\ 
   {\gamma \left( {{l_{1K}}} \right)} & {\gamma \left( {{l_{2K}}} \right)} &  \cdots  & {{\gamma }\left( 0 \right)}  \\ 
\end{array} } \right],
\end{align}
where ${l_{ij}} = L_{mn}^{\left( i \right)} - L_{mn}^{\left( j \right)}$ and $\gamma \left( {{l_{ij}}} \right)$ is the waveform auto-correlation function, i.e.,
\begin{align}
  \gamma \left( {{l_{ij}}} \right) & = {\mathbf{s}}_m^H{\left( {C_{L_{mn}^{\left( i \right)}}^T} \right)^H}C_{L_{mn}^{\left( j \right)}}^T{{\mathbf{s}}_m} \hfill  \nonumber\\
   &= {\mathbf{s}}_m^H{{\mathbf{J}}_{{l_{ij}}}}{{\mathbf{s}}_m} \hfill \nonumber \\
   &= \sum\limits_{k = {l_{ij}} + 1}^N {{s_m}\left( k \right)s_m^*\left( {k - {l_{ij}}} \right)}. 
\end{align}
Here, ${{\mathbf{J}}_n}$ is a shifting matrix \cite{he2009designing}, defined as 
\begin{align}
    {{\mathbf{J}}_n} = {\left[ {\begin{array}{*{20}{c}}
   {\overbrace {0 \cdots 0}^n1} & {} & 0  \\ 
   {} &  \ddots  & {}  \\ 
   {} & {} & 1  \\ 
   0 & {} & {}  \\ 
\end{array} } \right]_{N \times N}}.
\end{align}
Thus, 
\begin{align}
    \tau  = \frac{{{\text{tr}}\left( {\mathbf{\Phi }} \right)}}{K} &= \gamma \left( 0 \right), \\
    {\text{tr}}\left( {{{\mathbf{\Phi }}^2}} \right) &= K\left( {{{\left| {\gamma \left( 0 \right)} \right|}^2} + \sum\limits_{j \ne i}^K {\sum\limits_{i = 1}^K {{{\left| {\gamma \left( {{l_{ij}}} \right)} \right|}^2}} } } \right).
\end{align}
Then, according to Theorem  \ref{theorem_eigenvalue}, 
\begin{align}
    {\lambda _{\min }}\left( {\mathbf{\Phi }} \right) \ge \gamma \left( 0 \right) - \sqrt {K - 1} \sqrt {\sum\limits_{j \ne i}^K {\sum\limits_{i = 1}^K {{{\left| {\gamma \left( {{l_{ij}}} \right)} \right|}^2}} } } .
\end{align}
For unimodular sequence, it is easy to verify that 
\begin{align}
{\gamma \left( 0 \right)}  = N, \quad {\mathop {\sup }\limits_{i \in {\mathbb{N}}_{N + L_{mn}^{\left( {\max } \right)}}^ + } {{\left\| {{{\left( {{\mathbf{\Gamma }}_{mn}^T} \right)}^{\left( i \right)}}} \right\|}^2}}  = K.
\end{align}
We have 
\begin{align}
  \mu \left( V \right) &\le \frac{{N + L_{mn}^{\left( {\max } \right)}}}{K}\frac{{\mathop {\sup }\limits_{i \in {\mathbb{N}}_{N + L_{mn}^{\left( {\max } \right)}}^ + } {{\left\| {{{\left( {{\mathbf{\Gamma }}_{mn}^T} \right)}^{\left( i \right)}}} \right\|}^2}}}{{\gamma \left( 0 \right) - \sqrt {K - 1} \sqrt {\sum\limits_{j \ne i}^K {\sum\limits_{i = 1}^K {{{\left| {\gamma \left( {{l_{ij}}} \right)} \right|}^2}} } } }} \hfill \nonumber\\
   &= \frac{{N + L_{mn}^{\left( {\max } \right)}}}{{N - \sqrt {K - 1} \sqrt {\sum\limits_{j \ne i}^K {\sum\limits_{i = 1}^K {{{\left| {\gamma \left( {{l_{ij}}} \right)} \right|}^2}} } } }}. 
\end{align}
If the unimodular waveform sequences are designed to have ideal auto-correlation properties, i.e.,
\begin{align}
\gamma \left( l \right) = 0, \quad l = 1, \ldots ,\mathop {\max }\limits_{j \ne i} \left| {L_{mn}^{\left( i \right)} - L_{mn}^{\left( j \right)}} \right|,
\label{gamma coherent condition}
\end{align}
the coherence of the column space of ${{\bf{Z}}_{mn}}$ satisfies 
\begin{align}
    \mu \left( V \right) \le 1 + \frac{{L_{mn}^{\left( {\max } \right)}}}{N}.
\label{mu coherent condition}
\end{align}
\end{IEEEproof}

{\bf Remarks:} Theorem 4 suggests that the maximum time-difference-of-arrival, denoted as ${L_{mn}^{\left( {\max } \right)}}$, or alternatively the distribution of transmitting and receiving antennas, can influence the coherence of the radar data matrix. This implies that, given identical target locations, the recovery performance of matrix completion may vary depending on the geometry of the antenna setup.


\subsection{Ambiguity Function of WS-MIMO Radar}
\label{sucsec:af}
The ambiguity function (AF) characterizes radar's ability to distinguish two closely-spaced targets \cite{pinilla2022phase,radmard2014ambiguity, rendas1998ambiguity,ilioudis2016ambiguity}. In \cite{radmard2014ambiguity}, WS-MIMO radar AF is based on the ML and Kullback-directed divergence (KDD) \cite{rendas1998ambiguity}. 
Alternatively, \cite{ilioudis2016ambiguity} proposes an AF for distributed MIMO radar while avoiding the large matrix inversions. 
We adopt this definition of AF to evaluate the performance of WS-MIMO radar with different antenna geometries and SNRs. 
Recall the received signal 
\begin{align}
{y_{mn}}\left( t \right) =&\sum\limits_{q = 0}^{Q-1} \sqrt E \beta _{mn} {s_m}\left( {t - \tau _{mn} - qT_{\textrm{PRI}}} \right) \nonumber \\
& \times {e^{\mathrm{j}2\pi f_{mn}t}}{e^{ - j2\pi  {f_c} {\tau _{mn}}}}  + {w_{mn}}\left( t \right). \label{eq_re_sg} 
\end{align}
To further simplify (\ref{eq_re_sg}), define
\begin{align}
{\alpha_{mn}}\left( \bm{\theta} \right) =& \sqrt E \beta _{mn} \nonumber  {e^{ - j2\pi  {f_c} {\tau _{mn}}}},
\end{align}
and
\begin{align}
{\gamma_{mn}}\left( t, \bm{\theta}  \right) =\sum\limits_{q = 0}^{Q-1} {s_m}\left( {t - \tau _{mn} - qT_{\textrm{PRI}}} \right) {e^{\mathrm{j}2\pi f_{mn}t}},
\end{align}
where $\bm{\theta} = \left [ x, y, v_{x} , v_{y} \right ]^{\mathrm{T} } $ is the vector containing target position and Doppler velocity. Then (\ref{eq_re_sg}) becomes
\begin{align}
{y_{mn}}\left( t \right) ={\gamma_{mn}}\left( t, \bm{\theta}  \right) {\alpha_{mn}}\left( \bm{\theta} \right) + {w_{mn}}\left( t \right). \label{eq_re_sg_rewrite}
\end{align}
After sampling, we rewrite the discretized (\ref{eq_re_sg_rewrite}) as a $\left( {N + L_{mn}^{(\max)}} \right) \times 1 $ vector 
\begin{align}
{\mathbf{y}_{n,m} } ={\bm{\gamma}_{m}}\left( \bm{\theta}  , n\right) {{\alpha}_{n,m}}\left( \bm{\theta} \right) + {\mathbf{w}_{n}}. 
\end{align}
Collecting the samples for all antennas, we obtain the received signal matrix 
\begin{align}
\mathbf{Y}=\bm{\Upsilon} \left ( \bm{\theta}  \right ) \bm{\alpha} \left ( \bm{\theta}  \right ) +\mathbf{W},\label{complete_MIMO_eq}
\end{align}
which  turns out to be 
a $ \left ( N+L_{mn}^{(\max)} \right )M_{r} \times M_{r} $ block matrix 
\begin{align}
\mathbf{Y}= \begin{bmatrix}
  \bm{\gamma}_{1}&  0&  ...&0 \\
  0&  \bm{\gamma}_{2}& ... & 0 \\
  \vdots &  \vdots&  \ddots & \vdots \\
  0& 0 & ... & \bm{\gamma}_{M_{r}}
\end{bmatrix},
\end{align}
such that
\begin{align}
\bm{\gamma}_{n} &= \sum_{m=1}^{M_{t} } \mathbf{y}_{n,m},\;n = 1, 2, \cdots, M_{r}
\notag
\\ &= \sum_{m=1}^{M_{t}} {\bm{\gamma}_{m}}\left( \bm{\theta}  , n\right) {{\alpha}_{n,m}}\left( \bm{\theta} \right) + {\mathbf{w}_{n}}\;
\notag
\\ &= \bm{\Upsilon} \left ( \bm{\theta}, n  \right ) \bm{\alpha} \left ( \bm{\theta}, n  \right ) + \mathbf{w}_{n}.
\end{align}
where 
\begin{align}
 \bm{\Upsilon} \left ( \bm{\theta}, n  \right ) = \left [  \bm{\gamma}_{1}\left( \bm{\theta}  , n\right), \bm{\gamma}_{2}\left( \bm{\theta}, n\right),...,\bm{\gamma}_{M_{t}}\left( \bm{\theta}  , n\right) \right ] \in \mathbb{C}^{\left( {N + L_{mn}^{(\max)}} \right) \times M_{t} },
\end{align}

\begin{align}
 \bm{\alpha} \left ( \bm{\theta}, n  \right ) = \left [ {{\alpha}_{n,1}}\left( \bm{\theta} \right),{{\alpha}_{n,2}}\left( \bm{\theta} \right),...,{{\alpha}_{n,M_{t}}}\left( \bm{\theta} \right)  \right ]^{\mathrm{T} }   \in \mathbb{C}^{M_{t} \times 1}.
\end{align}
Then $\bm{\Upsilon} \left ( \bm{\theta}  \right )$ and $\bm{\alpha} \left ( \bm{\theta}  \right )$ in (\ref{complete_MIMO_eq}) are 
\begin{align}
 \bm{\Upsilon} \left ( \bm{\theta} \right ) = \begin{bmatrix}
  \bm{\Upsilon} \left ( \bm{\theta}, 1 \right )&  0&  ...&0 \\
  0&  \bm{\Upsilon} \left ( \bm{\theta}, 2 \right )& ... & 0 \\
  \vdots &  \vdots&  \ddots & \vdots \\
  0& 0 & ... & \bm{\Upsilon} \left ( \bm{\theta}, M_{r} \right )
\end{bmatrix}, 
\end{align}
and
\begin{align}
\bm{\alpha} \left ( \bm{\theta}  \right ) = \begin{bmatrix}
  \bm{\alpha} \left ( \bm{\theta}, 1  \right )&  0&  ...&0 \\
  0&  \bm{\alpha} \left ( \bm{\theta}, 2  \right )& ... & 0 \\
  \vdots &  \vdots&  \ddots & \vdots \\
  0& 0 & ... & \bm{\alpha} \left ( \bm{\theta}, M_{t}  \right )
\end{bmatrix}.  
\end{align}
According to \cite{ilioudis2016ambiguity} WS-MIMO AF is defined as
\begin{align}
 \mathcal{F} \left ( \bm{\theta}_{0}  ,\bm{\theta}  \right ) = 1 - \frac{I\left ( \bm{\theta}_{0};\bm{\theta} \right ) }{\sup I\left ( \bm{\theta}_{0};\bm{\theta} \right ) }.\label{WS_MIMO_AF_original}
\end{align}
where 
\begin{align}
I\left ( \bm{\theta}_{0};\bm{\theta} \right ) = \frac{1}{2}\left [ \mathrm{tr}\left [ \bm{R}_{\bm{\theta}}^{-1}\bm{R}_{\bm{\theta}}-\left( {N + L_{mn}^{(\max)}} \right) \times M_{r} \right ] -\mathrm{ln} \left | \bm{R}_{\bm{\theta}_{0}}^{-1}\bm{R}_{\bm{\theta}} \right |\right ].\label{KKD_definition}
\end{align}
is the KDD between two covariance matrices $\bm{R}_{\bm{\theta}_{0}}$ and $\bm{R}_{\bm{\theta}}$ with respect to received signal and the covariance matrix is
\begin{align}
\bm{R}_{\bm{\theta}} & =  E\left \{ \mathbf{Y}\mathbf{Y}^{H} \right \} \nonumber\\
& = E\left \{ \left ( \bm{\Upsilon} \left ( \bm{\theta}  \right ) \bm{\alpha} \left ( \bm{\theta}  \right ) +\mathbf{W} \right )\left ( \bm{\Upsilon} \left ( \bm{\theta}  \right ) \bm{\alpha} \left ( \bm{\theta}  \right ) +\mathbf{W} \right )^{H} \right \} \nonumber\\
& =  \bm{\Upsilon} \left ( \bm{\theta}\right )E\left\{\bm{\alpha} \left ( \bm{\theta}  \right )\bm{\alpha} \left ( \bm{\theta}  \right )^{H} \right\}\bm{\Upsilon} \left ( \bm{\theta}\right )^{H} + \bm{\sigma}_{n}^{2}\mathbf{I}\nonumber\\
& = \bm{\Upsilon} \left ( \bm{\theta}\right )\bm{C}\left(\bm{\theta}\right)\bm{\Upsilon} \left ( \bm{\theta}\right )^{H} + \bm{\sigma}_{n}^{2}\mathbf{I}.\label{covirance_matrix}
\end{align}
Substituting (\ref{covirance_matrix}) into (\ref{KKD_definition}) and applying the constant energy and SNR conditions \cite{ilioudis2016ambiguity}, we obtain the AF as
\begin{align}
 \mathcal{F} \left ( \bm{\theta}_{0}  ,\bm{\theta}  \right ) =\frac{1}{M_{t}M_{r}} \mathrm{tr} \left [ \left | \bm{\Upsilon} \left ( \bm{\theta}_{0} \right )\bm{\Upsilon}^{\mathrm{H} } \left ( \bm{\theta} \right ) \right | ^{2} \right ].\label{WS_MIMO_AF}
\end{align}
When $\theta=\theta_{0}$, the KDD $ I\left ( \bm{\theta}_{0};\bm{\theta} \right ) $ reached its minimum and then ambiguity function $ \mathcal{F} \left ( \bm{\theta}_{0}  ,\bm{\theta}  \right )$ achieves 
 its maximum. Unlike the AF of a monostatic radar systems\cite{levanon2004radar}, the WS-MIMO AF introduced in (\ref{WS_MIMO_AF}) includes the impact of the geometry of antenna distribution on the performance of WS-MIMO systems. This is helpful in determining an appropriate antenna configuration for real applications.

\section{Numerical Experiments}
\label{sec:num_exp}

We evaluated the performance of our proposed MC-WS-MIMO radar through numerical experiments. 
Throughout all experiments, we employed singular value thresholding (SVT) \cite{cai2010svt} algorithm at the fusion center to recover the data matrix ${{{\bf{Z}}_{mn}}}$ corresponding to the $m$-th-Tx-and-$n$-th-Rx pair from its partial samples ${{{\bf{X}}_{mn}}}$.

\begin{figure*}
\centering
 \subfloat[][]{\includegraphics[height=1.4 in ]{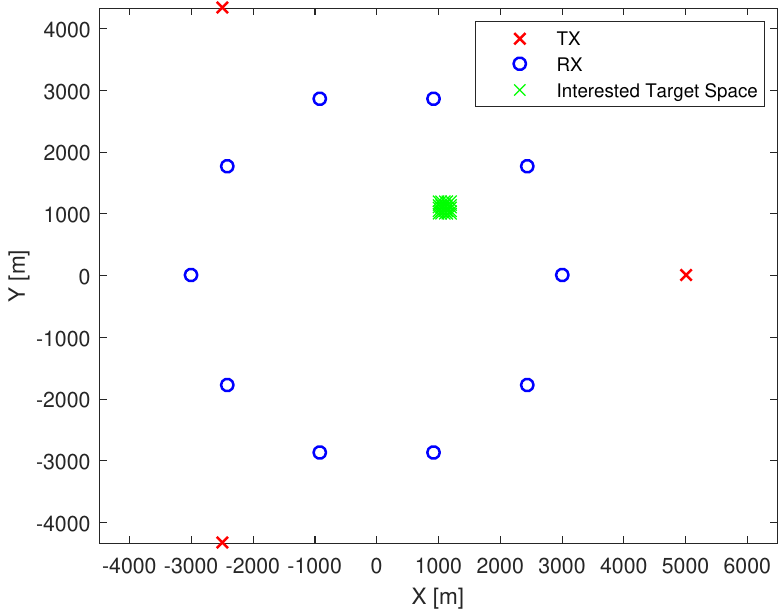}}
 \subfloat[][]{\includegraphics[height=1.4 in]{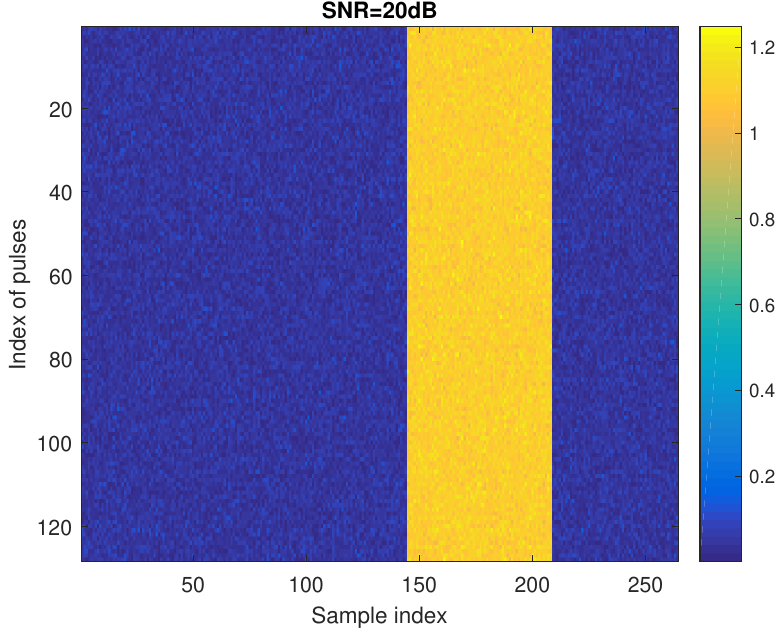}}
 \subfloat[][]{\includegraphics[height=1.4 in]{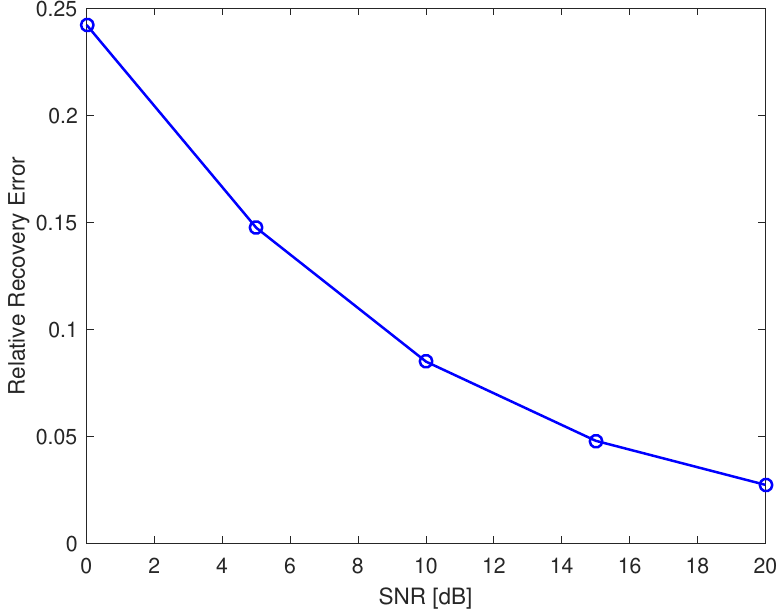}}
 \subfloat[][]{\includegraphics[height=1.4 in]{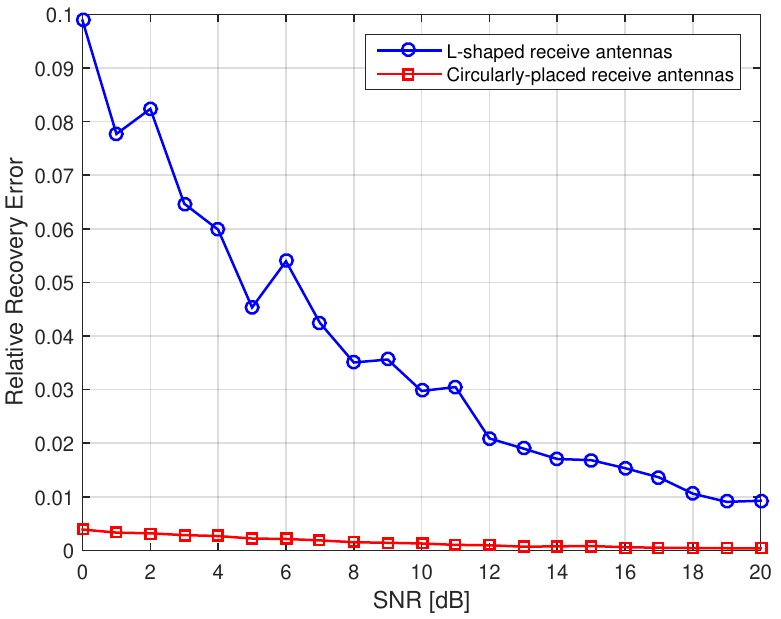}}
  \caption{(a) An illustration of WS-MIMO radar with $M_t=3$ transmit and $M_r=10$ receive  antennas. The transmit and receive antennas are uniformly distributed on circles of radii $5000$ m and $3000$ m, respectively. The targets of interest are distributed within the area $\left[ {1000,1200} \right] \times \left[ {1000,1200} \right]$. The distance dimensions are in meters (m); (b) The data matrix at receive antenna ${{\bf{p}}_r} = \left[ {2427.1, - 1763.4} \right]^T$ corresponding to signal from the transmitter at ${{\bf{p}}_t} = \left[ { - 2500, - 4330.1} \right]^T$ reflected from the target located at ${{\bf{p}}} = \left[ {1100,1100} \right]^T$. The distance dimensions are in meters. The length of sampling window and SNR were set at $N+ L_{\max}=264$ and $20$ dB, respectively; (c) Matrix recovery error $\varepsilon$ as a function of SNR for the WS-MIMO configuration of Fig. \ref{fig_WS_example_MC_error} (a). The error is averaged over all $30$ Tx-Rx pairs; (d) Matrix recovery error $\varepsilon$ as a function of SNR for the WS-MIMO configuration of Fig. \ref{fig:array_config_ambiguity_SNR_20dB} (a) and \ref{fig:array_config_ambiguity_SNR_20dB} (d) respectively. The blue line represents the recovery error versus SNR for the L-shaped receive antennas configuration, and the red line represents the recovery error versus SNR for the Circularly-placed receive antennas configuration. The error is averaged over all $30$ Tx-Rx pairs. }
  \label{fig_WS_example_MC_error}
\end{figure*}

\subsection{Reconstruction Error under Different Antenna Geometry}

We considered a WS-MIMO radar with $M_t=3$ transmit and $M_r=10$ receive antennas (Fig.~\ref{fig_WS_example_MC_error} (a)) uniformly distributed over circles with radii $5000$ m and $3000$ m, respectively. The targets of interest are distributed in the area $\mathcal S=\left[ {1000,1200} \right] \times \left[ {1000,1200} \right]$ m$^2$. The transmitters emit Hadamard sequences \cite{proakis2001digital} of length $N=64$. The rows of the Hadamard matrix are mutually orthogonal to each other and can be used as Walsh codes in a MIMO radar. The carrier frequency parameters were set to $f_0 =5$ GHz and $\Delta f = 50$ MHz. The CPI comprised of $Q=128$ pulses with $T_{\rm PRI} = 25$ ms and $T_p=6.4$ $\mu$s.

In the Nyquist case, the sampling frequency at the receive antennas is $f_s = 10$ MHz. In order to 
unambiguously sample the area ${\mathcal S}$, we choose the length of sampling window as $N+{L_{\max}}=264$, where ${L_{\max }} = \mathop {\max }\limits_{m,n} \left\{ {L_{mn}^{(\max)}} \right\} = 200$. A single target located at ${{\bf{p}}} = \left[ {1100,1100} \right]^T$ m with velocity ${\bm{\nu }} = \left[ {10,10} \right]^T$ m/s is considered for recovery. 
Fig.~\ref{fig_WS_example_MC_error} (b) plots the data matrix ${\bf Z}_{mn}$ for the receive antenna ${{\bf{p}}_r} = \left[ {2427.1, - 1763.4} \right]^T$ m and the reflected echo for the transmitter at ${{\bf{p}}_t} = \left[ { - 2500, - 4330.1} \right]^T$ m at $\textrm{SNR}=20$ dB. The noise at each receive antenna is generated independently for different Tx-Rx antenna pairs. It follows from Fig.~\ref{fig_WS_example_MC_error} (b) that the data matrix is rank-$1$ and the samples of reflected echo start at range-sample index of $L_{mn}^{(1)}=144$.

In each CPI, the $n$-th receive antenna samples only $50$\%  of matrix ${{{\bf{Z}}_{mn}}}$, $m=1,\cdots,M_t$ uniformly at random. At the fusion center, when these matrices are completed using SVT, we characterize the recovery performance by relative error defined as $\varepsilon  = \frac{{{{\left\| {{\mathbf{Z}_{mn}} - {\mathbf{\hat Z}_{mn}}} \right\|}_{\mathcal{F}}}}}{{{{\left\| {\mathbf{Z}_{mn}} \right\|}_{\mathcal{F}} }}}$, where $ {\mathbf{\hat Z}_{mn}}$ denotes the recovered matrix. For different values of SNR,  Fig.~\ref{fig_WS_example_MC_error} (c) plots the recovery error averaged over all $M_r\times M_t=30$ Tx-Rx pairs. The error drops to approximately $3$\% at $\textrm{SNR}=20$ dB. Fig.~\ref{fig_WS_example_MC_error} (d) compares the recovery errors (averaged over 100 trials) w.r.t. two different antenna configurations.
It follows that the circularly antenna configuration generally exhibits an improved and robust recovery over the L-shaped geometry. 
Following equations (\ref{mu coherent condition}) and (\ref{gamma coherent condition}), any change in the placement of antennas results in a corresponding change in the value of $L_{mn}^{\left(max\right)}$, which may lead to a violation of the coherence condition and a deterioration of the matrix recovery performance. In Fig.~\ref{fig_WS_example_MC_error} (d), $L_{mn}^{\left(max\right)}$ is $200$ ($116$) for the circularly-placed (L-shaped) antennas. Therefore, the coherence condition is not adequately satisfied for L-shaped geometry, resulting in larger relative recovery errors. The outcomes of this simulation substantiate the validity of Theorem 4.

\begin{figure*}
\centering
 \subfloat[][]{\includegraphics[width=0.23\textwidth ]{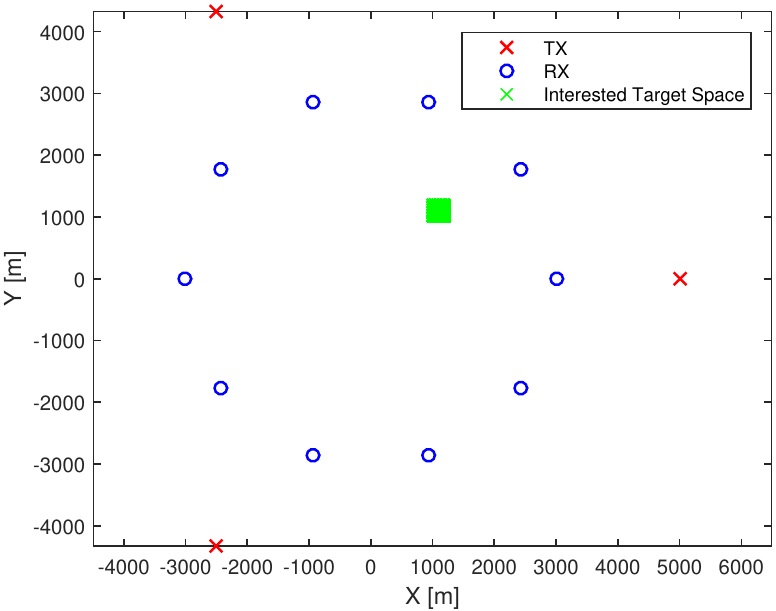}}
 \subfloat[][]{\includegraphics[width=0.25\textwidth]{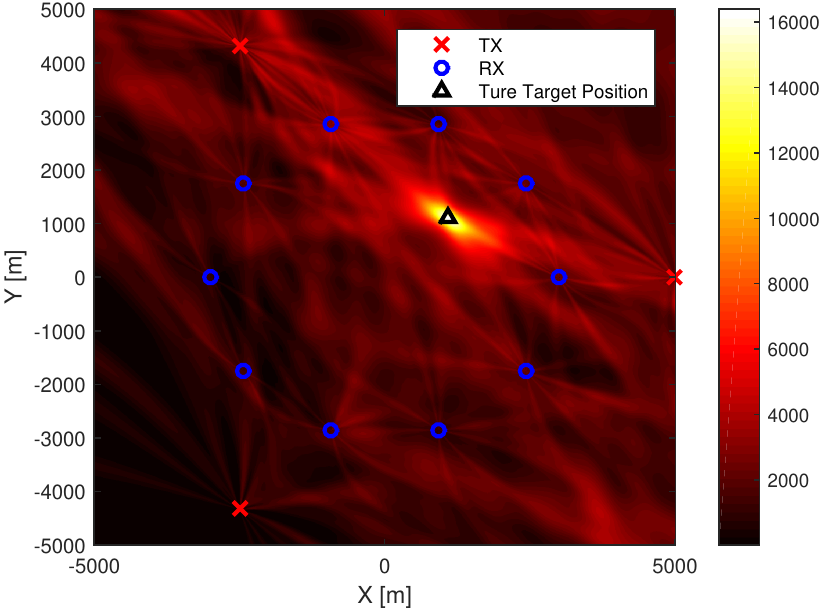}}
 \subfloat[][]{\includegraphics[width=0.265\textwidth]{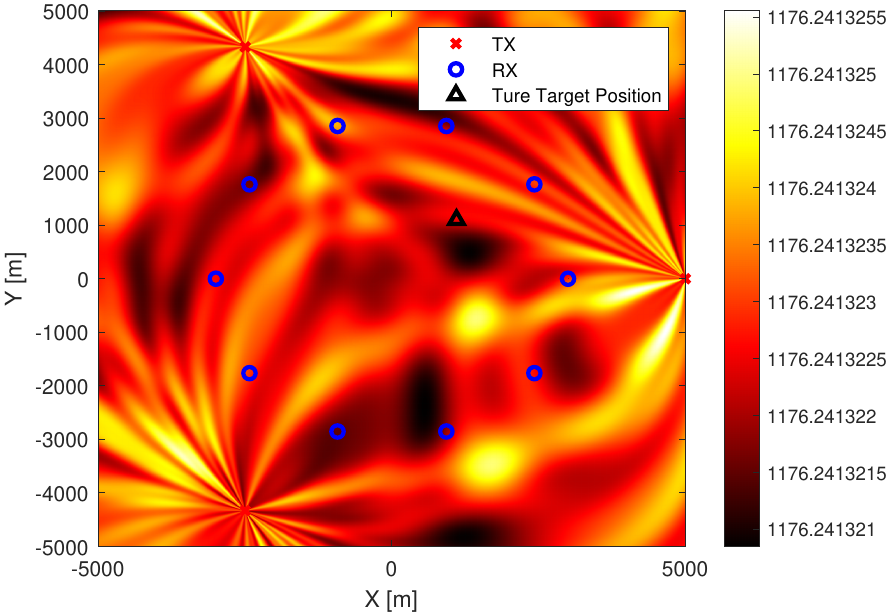}}\quad
 \subfloat[][]{\includegraphics[width=0.23\textwidth]{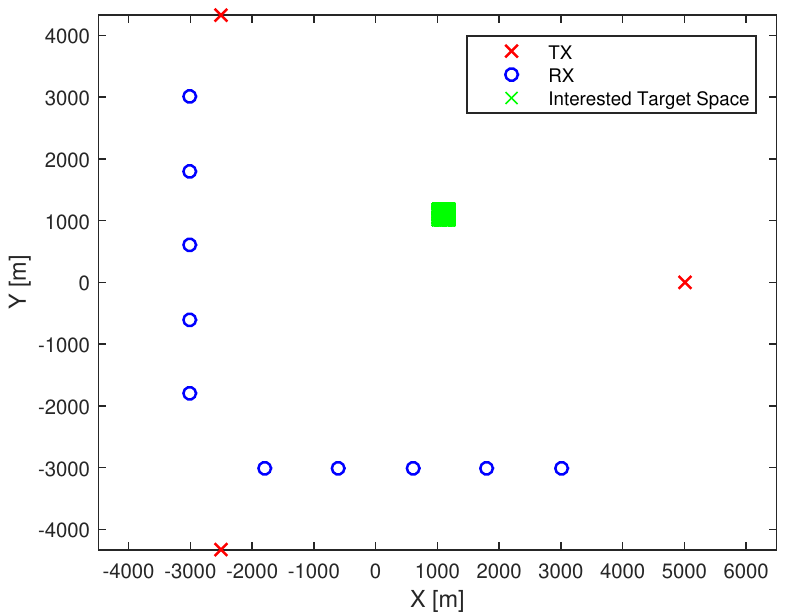}}
 \subfloat[][]{\includegraphics[width=0.25\textwidth]{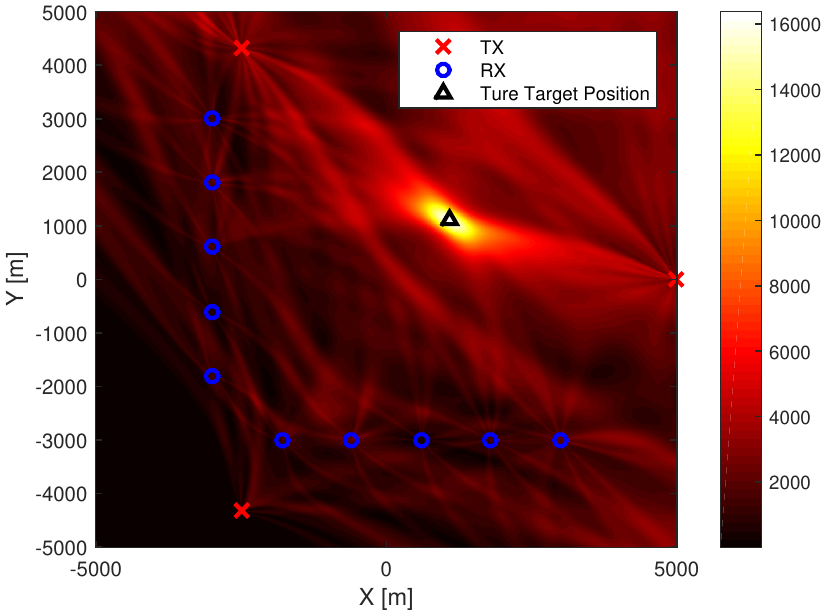}}
 \subfloat[][]{\includegraphics[width=0.265\textwidth]{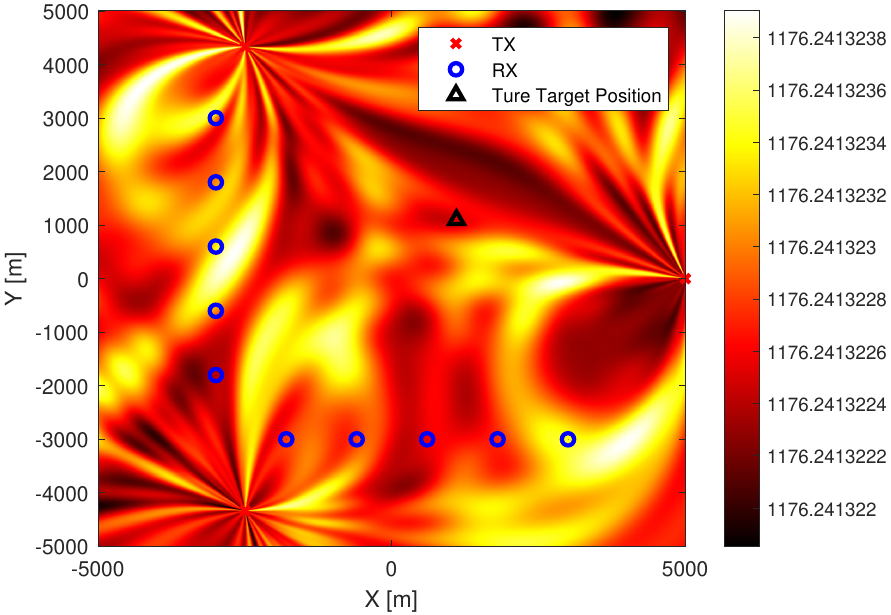}}\quad
 \subfloat[][]{\includegraphics[width=0.23\textwidth]{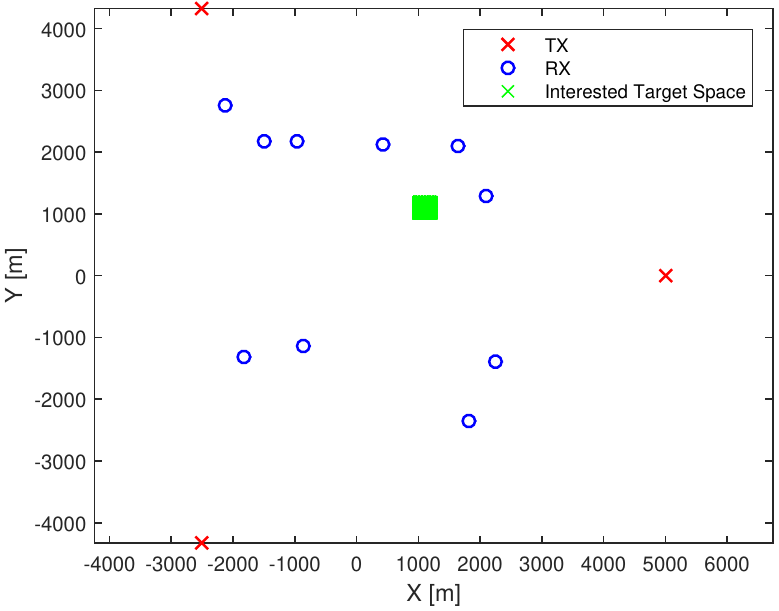}}
 \subfloat[][]{\includegraphics[width=0.25\textwidth]{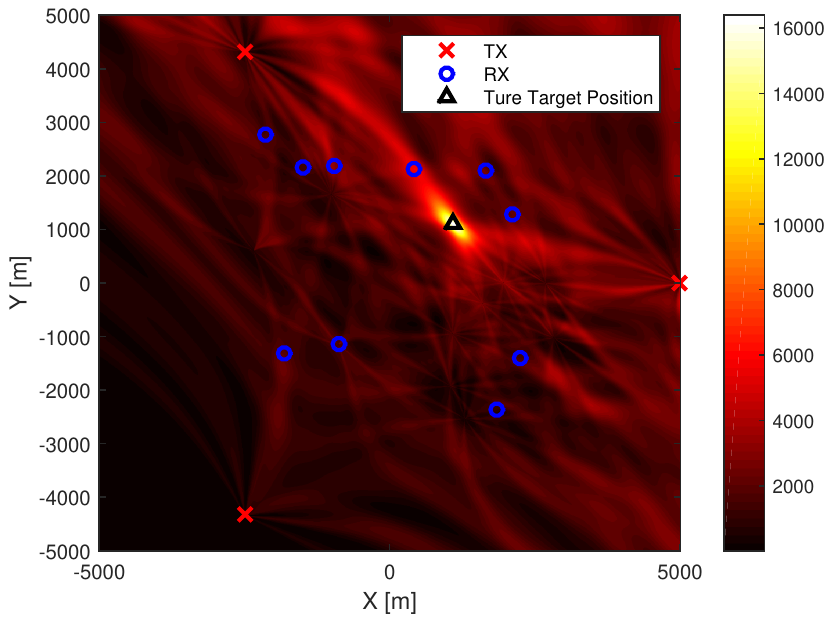}}
 \subfloat[][]{\includegraphics[width=0.265\textwidth]{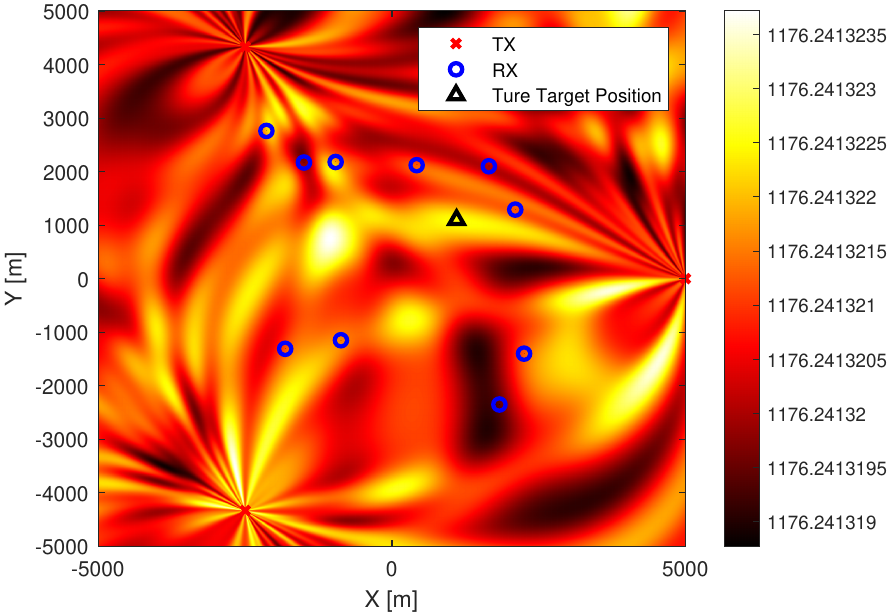}}
  \caption{The WS-MIMO radar antenna geometries and corresponding AF: (a) Circularly-placed receive antennas (b) AF corresponding to the configuration in (a) under $100$\% sampling rate. (c) AF corresponding to the configuration in (a) under $20$\% sampling rate. (d-f) As in (a)-(c), respectively but for L-shaped receive antennas. (g-i) As in (a)-(b), respectively but for randomly placed receive antennas.}
  \label{fig:array_config_ambiguity_SNR_20dB}
\end{figure*}


\subsection{Ambiguity Function Simulation Results}
The WS-MIMO radar waveform based on Hadamard codes is used in the simulation. Fig.~\ref{fig:array_config_ambiguity_SNR_20dB} shows the AF corresponding to different antenna geometry at SNR $=20$ dB. 
In this case, we still consider WS-MIMO radar with $M_t=3$ transmit and $M_r=10$ receive antennas. The 3 transmit antennas are uniformly distributed over a circle with radii $5000$ m as before but the 10 receive antennas distribution is changed to a circle with radii $3000$ m (Fig.~\ref{fig:array_config_ambiguity_SNR_20dB} (a)). Fig.~\ref{fig:array_config_ambiguity_SNR_20dB} (c) displays another geometry, wherein the receive antennas are linearly spaced in an L-shape. We also consider a random geometry in Fig.~\ref{fig:array_config_ambiguity_SNR_20dB}e, wherein the receive antennas are randomly distributed over $\left [ -3000\mathrm{m}, 3000\mathrm{m}   \right ] \times \left [ -3000\mathrm{m}, 3000\mathrm{m} \right ]$. 
From Fig.~\ref{fig:array_config_ambiguity_SNR_20dB} (b), the AF achieves its maximum at the target's position which is consistent with the property of AF stated in Section IV.B. 
We observe that the AF corresponding to the L-shape linear distribution has a larger ambiguous range compared to the circular and random placement of the receive antennas. Figs.~\ref{fig:array_config_ambiguity_SNR_20dB} (c), (f), and (i) show the AFs corresponding to different array configurations under $20$\% sampling rate. It follows that the AF is degenerated under low sampling rate because stronger ambiguities appear at the positions of the transmit/receive antennas and the target. This indicates that the accuracy of localization also decreases with sub-sampling. In addition, the circularly-placed geometry has a better AF compared with the other configurations under sub-sampling.

\subsection{Localization Performance Comparison}

\subsubsection{Different number of antennas}

\begin{figure*}
\centering
  \subfloat[][]{\includegraphics[height=1.62 in ]{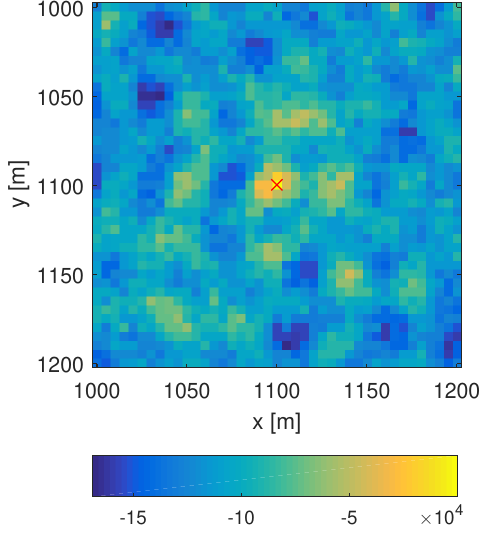}}
    \subfloat[][]{\includegraphics[height=1.62 in]{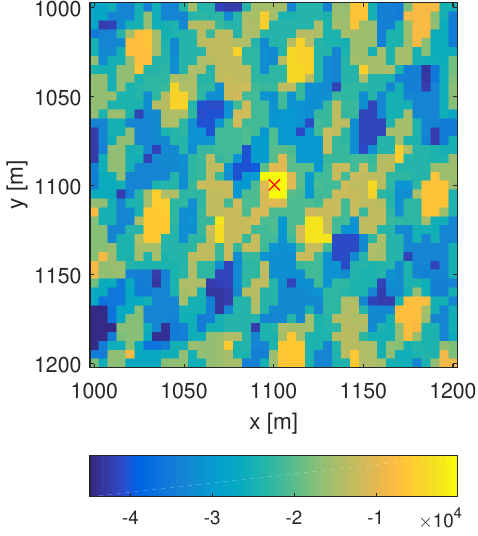}}
  \subfloat[][]{\includegraphics[height=1.62 in]{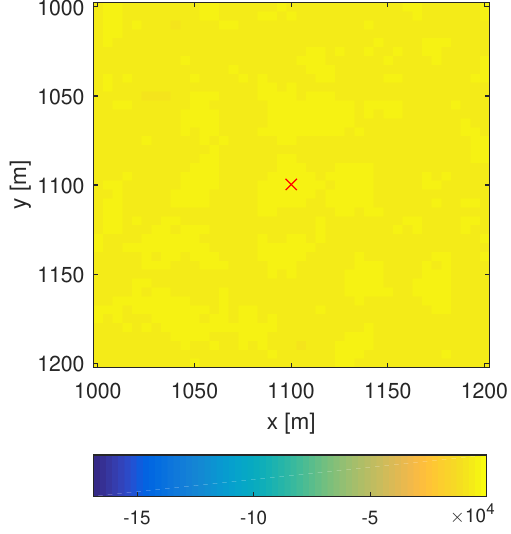}}
    \subfloat[][]{\includegraphics[height=1.62 in]{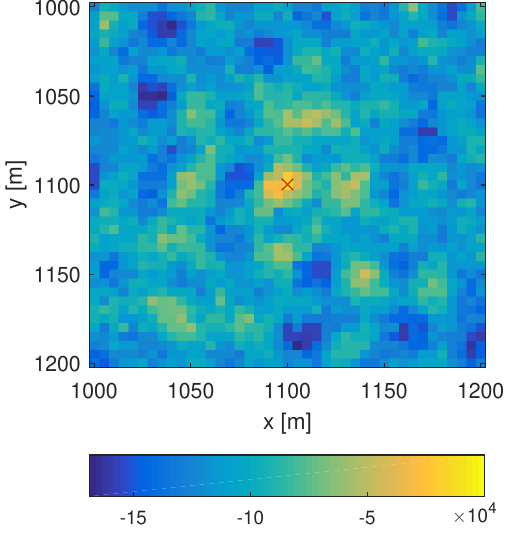}}

  \caption{The ML estimation result using subsampling and matrix recovery signals for a target at ${\bf{p}} = \left[ {1100,1100} \right]^T$ m with a velocity of ${\bm{\nu}} = \left[ {10,10} \right]^T$ m/s under the WS-MIMO radar configuration given in Fig.~\ref{fig_WS_example_MC_error} (a), SNR $=20$ dB. The red cross indicates the location of target. (a) target localization using ML under $M_t=3, M_r=10$ and $50$\% sampling rate; (b) target localization using ML under $M_t=2, M_r=4$ and $50$\% sampling rate;  (c) target localization using ML with $20$\%  subsampled signal; (d) target localization using ML after MC-based recovery. }
  \label{fig_comp_recovery_single}
\end{figure*}

Fig.~\ref{fig_comp_recovery_single} (a) plots the ML values for the 2-D search over the area $\mathcal S$ for a single target at ${\bf{p}} = \left[ {1100,1100} \right]^T$ m with velocity ${\bm{\nu}} = \left[ {10,10} \right]^T$ m/s at SNR of $20$ dB. 
Fig.~\ref{fig_comp_recovery_single} (a) shows that the ML estimate corresponds to the true location of the target. For comparison, in Fig.~\ref{fig_comp_recovery_single} (b), we also show the ML estimate for the same setting as in Fig.~\ref{fig_comp_recovery_single} (a) except that the number of antennas is reduced to $M_t=2$ and $M_r=4$. 
We note  the range resolution decreases with the number of transmit-receive pairs.

After reconstructing the matrices ${\bf{Z}}_{mn}$, we show the ML-based target location estimation. Fig.~\ref{fig_comp_recovery_single} (c) and (d) show the ML performance for WS-MIMO radar configuration with and without MC-based recovery. At SNR $=20$ dB and subsampling at $20$\% rate, when ML is applied directly on subsampled signal (Fig.~\ref{fig_comp_recovery_single} (c)), estimation with ML is quite inferior when compared with its application on MC-based recovery (Fig.~\ref{fig_comp_recovery_single} (d)) wherein the recovery error is around $\varepsilon=4.8$\%.

\subsubsection{Comparison with geometric method}

\begin{figure*}
\centering
 \subfloat[][]{\includegraphics[width=0.28\textwidth]{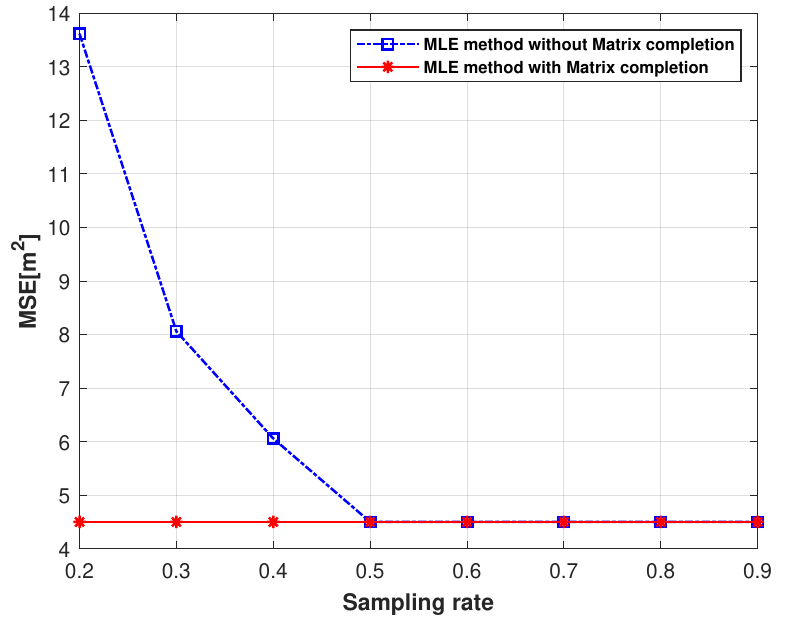}}
 \subfloat[][]{\includegraphics[width=0.28\textwidth]{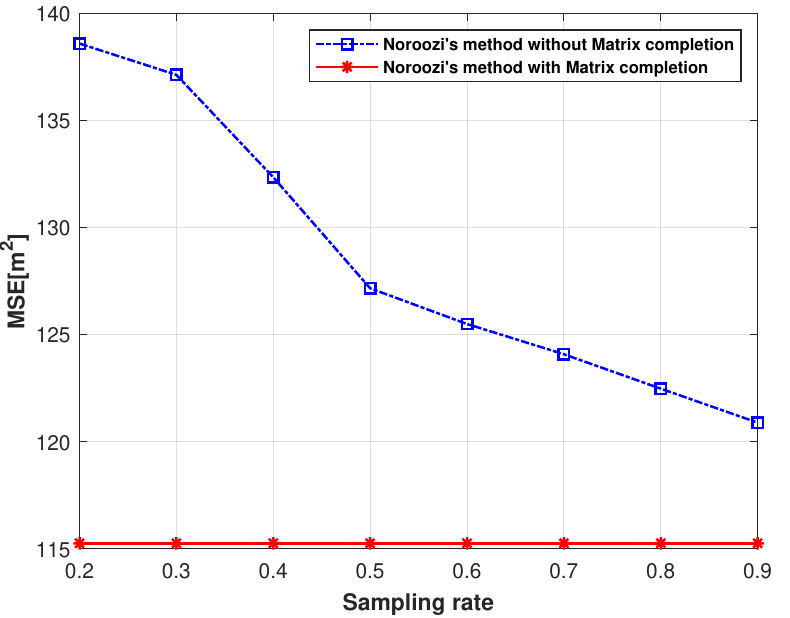}}
 \subfloat[][]{\includegraphics[width=0.28\textwidth]{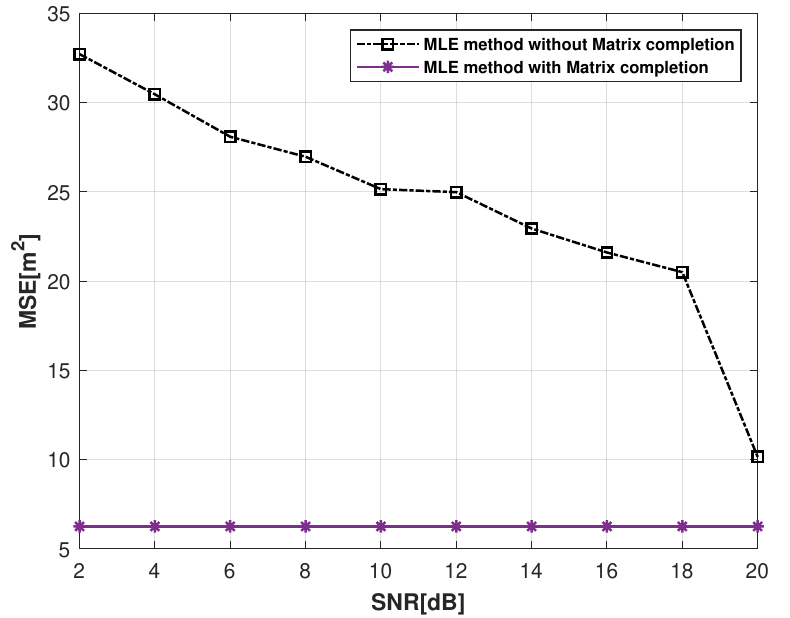}}\quad
 \subfloat[][]{\includegraphics[width=0.28\textwidth]{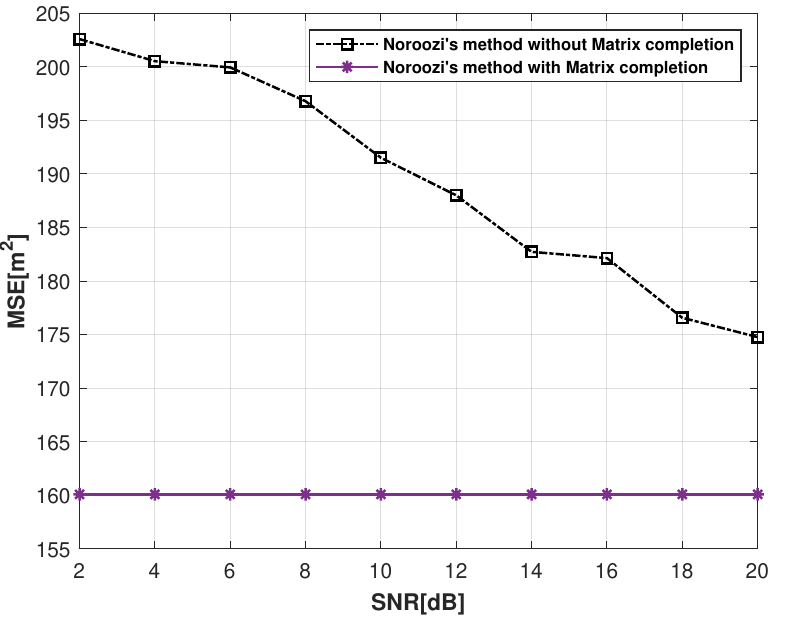}}
 \subfloat[][]{\includegraphics[width=0.28\textwidth]{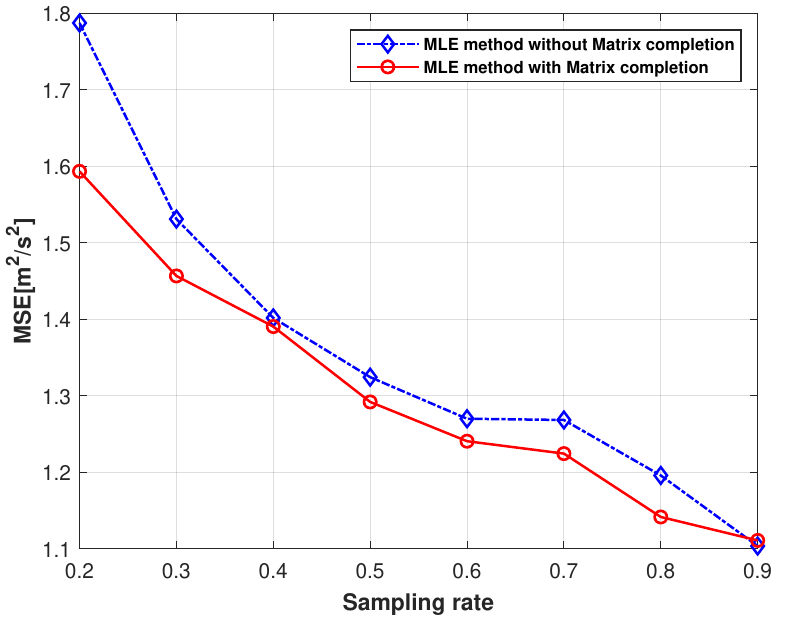}}
 \subfloat[][]{\includegraphics[width=0.28\textwidth]{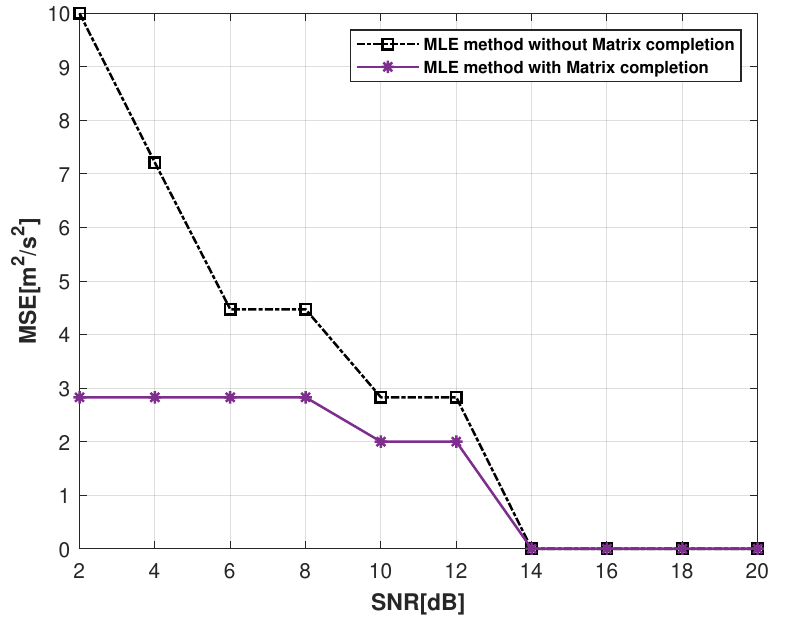}}
 
  \caption{Single target location and velocity estimation MSE using different methods under different sampling rates and SNRs, $M_t=3$, $M_r=10$. (a) Target location estimation with ML method  on sub-sampled signal and MC-based recovered signal under  SNR=20dB; (b) Target location estimation with geometric method \cite{noroozi2015target} on sub-sampled signal and MC-based recovered signal under  SNR=20dB; (c) Target location estimation with ML method  on sub-sampled signal and MC-based recovered signal under sampling rate of $20$\%; (d) Target location estimation with geometric method \cite{noroozi2015target} on sub-sampled signal and MC-based recovered signal under sampling rate of $20$\%; (e) Target velocity estimation with ML method on sub-sampled signal and MC-based recovered signal versus sampling rates; (f) Target velocity estimation with geometric method \cite{noroozi2015target} on sub-sampled signal and MC-based recovered signal versus SNRs. }
  \label{fig_comp_single_target}
\end{figure*}

Fig.~\ref{fig_comp_single_target} (a), (b) shows the  mean squared errors (MSEs) of single target localization estimation with ML and geometric \cite{noroozi2015target} methods while increasing samples from $20$\% to $90$\%. 
We compute the MSE of target location estimation as $ \frac{\sum_{i=1}^{K}\left (\hat{\bm{p}}^{\left ( i \right )}-\bm{p}^{\left ( i \right ) }\right )^{2}}{K} $.  The MSE of target velocity estimation can be calculated in a similar way.
In the simulation, the target was set at $\left[ {1100 \textrm{ m},1100 \textrm{ m}} \right]^T$ with velocity of $\left[ {10 \textrm{ m/s}, 10 \textrm{ m/s}} \right]^T$. The antenna distribution is the same as in Fig.~\ref{fig_WS_example_MC_error} (a). The MC performances of ML and geometric method do not change a lot with the number of samples, thereby demonstrating the robustness of MC as well as the redundancy (or low rankness) of data. 
Total $1000$ Monte Carlo experiments were conducted.
The MSEs of the single target localization estimation with ML and geometric methods versus different SNRs are shown in Fig.~\ref{fig_comp_single_target} (c), (d). The ML yields a smaller estimation MSE than the geometric method under different SNRs. Besides, the MSEs for all methods are nearly the same under different SNR values. Fig.~\ref{fig_comp_single_target} (e), (f) shows the single target velocity estimation MSE of the ML method under different sampling rates and SNRs. The velocity estimation is improved through MC when SNR is varied.

\vspace{-2mm}

\section{Summary}
\label{sec:summ}
We proposed the MC-WS-MIMO radar with FDM to detect spatially diverse targets. We showed that the received signal for each Tx-Rx pair over a CPI can be modeled as a low-rank data matrix. Reduced rate sampling of the signal at each receiver results in this matrix becoming partially observed. We retrieve its missing entries using MC methods. Despite sampling at low rates, our method retrieved the unknown off-grid target parameters. Our experiments indicate target parameter recovery with an accuracy of approximately $95$\% at $20$ dB SNR when the sampling rate is reduced to $20$\%. Our MC-based recovery is beneficial in enhancing the accuracy and robustnesss of target localization and velocity estimation in SNR-deficient scenarios. 
We show that further improvement in MC-based recovery is possible by analyzing the AFs for different antenna placements. This is meaningful for radar engineers in practical system design.

\bibliographystyle{IEEEtran}
\bibliography{main}

\end{document}